\UseRawInputEncoding
\documentclass[AMA,STIX1COL]{WileyNJD-v2}

\articletype{Article Type}%

\received{July 2024}
\revised{---}
\accepted{----}

\begin{document}

\title{Gauge fields and four interactions in the trigintaduonion spaces}

\author[1,2]{Zi-Hua Weng*}



\authormark{WENG}

\address[1]{\orgdiv{School of Aerospace Engineering}, \orgname{Xiamen University}, \orgaddress{\state{Xiamen 361005}, \country{China}}}

\address[2]{\orgdiv{College of Physical Science and Technology}, \orgname{Xiamen University}, \orgaddress{\state{Xiamen 361005}, \country{China}}}


\corres{*School of Aerospace Engineering, Xiamen University, Xiamen 361005, China. \email{xmuwzh@xmu.edu.cn}}


\abstract[Summary]{
The paper aims to apply the trigintaduonion spaces to explore the physical properties of four interactions simultaneously, including the electromagnetic fields, gravitational fields, weak nuclear fields, and strong nuclear fields. J. C. Maxwell first applied the algebra of quaternions to study the physical properties of electromagnetic fields. It inspired some subsequent scholars to introduce the quaternions, octonions, sedenions, and trigintaduonions to research the electromagnetic fields, gravitational fields, weak nuclear fields, strong nuclear fields, quantum mechanics, gauge fields, and curved spaces and so forth. The algebra of trigintaduonions is able to discuss the physical quantities of four interactions, including the field potential, field strength, field source, linear momentum, angular momentum, torque, and force. In the field theories described with the algebra of trigintaduonions, the weak nuclear field is composed of three types of fundamental fields. These three fundamental fields, related to weak nuclear fields, can describe the physical properties of weak nuclear fields collectively. This is consistent with the conclusion of the electroweak theory. Meanwhile the strong nuclear field consists of three types of fundamental fields. These three fundamental fields relevant to strong nuclear fields may investigate the physical properties of strong nuclear fields mutually. It is coincident with the deduction of quark theory. According to the properties of trigintaduonions, one can deduce the Yang-Mills equation related to the gauge fields. It means that the electromagnetic field occupies a quaternion space. The gravitational field owns one different quaternion space. The weak nuclear fields occupy three mutually independent quaternion spaces. The properties of weak nuclear fields are different from those of electromagnetic fields or gravitational fields. According to the multiplication table of trigintaduonion spaces, the strong nuclear fields own three conjugate quaternion spaces independent of each other. These explorations further deepen the understanding of the physical properties of weak and strong nuclear fields.
}


\jnlcitation{\cname{%
\author{Weng Z.-H.},
} (\cyear{2024}),
\ctitle{Gauge fields and four interactions in the trigintaduonion spaces},
\cjournal{Math Meth Appl Sci.},
\cvol{2024;00:1--6}.}

\maketitle


\section{\label{sec:level1}Introduction}

Is it possible to apply one single four-dimensional space to explore simultaneously the physical properties of electromagnetic fields, gravitational fields, weak nuclear fields, and strong nuclear fields? Must the space for weak nuclear fields be the same as that for electromagnetic fields? Why should three gauge fields be utilized to discuss the physical properties of weak nuclear fields? How many gauge fields are needed to study the physical properties of a strong nuclear field? For a long time, these difficulties have been troubling and tormenting many scholars. Over the years, these challenges have prompted some scholars to propose a few hypotheses in an attempt to solve these problems. Until recently, the emergence of field theory described by the trigintaduonions \cite{saini} (trigintaduonion field theory, for short) answered these puzzles partially. The trigintaduonion field theory claims to be able to achieve some equations of electromagnetic fields, gravitational fields, weak nuclear fields, and strong nuclear fields simultaneously, including the Yang-Mills equation relevant to weak nuclear fields.

In 1954, C. N. Yang and R. Mills extended the concepts of quantum electrodynamics to the non-Abelian groups to attempt to explain the strong interactions. In 1960, J. Goldstone, Y. Nambu, and G. Jona-Lasinio proposed the concept of particles obtaining mass through the spontaneous breaking of symmetry in the massless theory. These efforts have led to the successful application of Yang-Mills equation to the electroweak theory and quark theory.

In 1968, S. Weinberg and A. Salam established the electroweak theory, on the basis of the electroweak unified model of S. L. Glashow. The theory believes that the electromagnetic interaction and weak interaction belong to a unified interaction. The seemingly unrelated electromagnetic field and weak nuclear field are unified in this theory organically. It is necessary to introduce four gauge fields in the unified theory. In the following decades, the electroweak theory withstood numerous high-energy physics experiments.

The scholars believe that there exists the strong interaction between two hadrons. However the strong interaction is quite distinct from the weak interaction. The strong interaction is much more complicated than the weak interaction, and its interaction intensity is much stronger. In 1964, M. Gell-Mann and G. Zweig independently proposed that the hadrons such as the neutrons and protons are composed of quarks. Many predictions derived from the quark theory have been experimentally confirmed. Moreover, the scholars also put forward several other types of hypotheses, attempting to explain some physical phenomena of strong interactions.

The existing field theories have achieved many achievements, including the electroweak theory and quark theory. However, there are some puzzles derived from the existing field theories. After comparison, it can be found that the existing research has three problems as follows.

(1) Single four-dimensional space. The existing field theories are incapable of describing simultaneously the physical properties of electromagnetic fields, gravitational fields, weak nuclear fields, and strong nuclear fields. A single four-dimensional space is an obsolete physical concept and research pitfall. The existing field theories advocate researching the physical properties of these four interactions simultaneously in a single four-dimensional space. Attempting to achieve this goal is quite arduous, and even misguided. This self imposed and erroneous cognition limits the applicability of existing field theories.

(2) Three gauge fields. The existing field theories suggest that the weak nuclear field is a fundamental field. However three gauge fields are needed to explore the physical properties of weak nuclear fields. This is a contradiction in the narrative. If the weak nuclear field is a fundamental field, only one fundamental field is needed to discuss the physical properties of one weak nuclear field. On the contrary, if it is necessary to apply three gauge fields to depict the physical properties of a weak nuclear field, the weak nuclear field is not a fundamental field. This indicates that the existing field theories have become conceptually chaotic and inadequate.

(3) Narrative contradiction. In the existing field theories, the strong nuclear field is one type of fundamental field, but three gauge fields are required to study the physical properties of strong nuclear fields. This appears chaotic in the narrative. If the strong nuclear field is a fundamental field, only one fundamental field is needed to investigate the physical properties of strong nuclear fields. On the contrary, the strong nuclear field is not a fundamental field, if we have to apply three gauge fields to depict the physical properties of a strong nuclear field. This indicates that the existing field theories have become conceptually vague and lacks skills.

In contrast, several puzzles derived from the existing field theories can be solved in the trigintaduonion field theory, improved the unified theory related to the existing four interactions to a certain extent. J. C. Maxwell first applied the algebra of quaternions to explore the physical properties of electromagnetic fields. This inspired subsequent scholars to apply the quaternions \cite{rawat}, octonions \cite{moffat,tanisli1}, sedenions \cite{demir1,mironov2}, and trigintaduonions \cite{cariow,gul} to investigate the physical properties of some fundamental fields. Some scholars utilize the algebra of quaternions to research the Dirac wave equations \cite{sandhya,mironov3}, electromagnetic equations \cite{mironov1}, quantum mechanics \cite{deleo,bernevig}, gravitational theory, dark matters, dyonic matters \cite{pathak1,pathak2}, and curved spaces and so on. Several scholars apply the algebra of octonions to explore the Dirac wave equations \cite{morita}, curved spaces \cite{bossard}, electromagnetic equations \cite{chanyal3,chanyal4}, gravitational theory, dark matters, strong interactions \cite{furui,chanyal2}, Yang-Mills equations \cite{majid,farrill}, and Aharonov-Bohm effect \cite{a-b,tonomura}, plasma \cite{demir3}, and phase factors \cite{berry} and others. A few scholars introduce the algebra of sedenions \cite{demir2} to discuss the Dirac wave equation, gravitational equations, electromagnetic equations, curved spaces, and dark matters and so on.

However, the fundamental fields and adjoint fields in the sedenion spaces are unable to resolve some physical properties of weak nuclear fields. By contrast, the fundamental fields and adjoint fields in the trigintaduonion spaces are capable of describing several physical properties of weak nuclear fields.

Through comparison and analysis, we can find some remarkable characteristics of applying trigintaduonions to study four types of interactions.

(1) Multiple quaternion spaces. In the trigintaduonion spaces, it is capable of exploring the physical properties of electromagnetic fields, gravitational fields, weak nuclear fields, and strong nuclear fields simultaneously. The trigintaduonion space is one appropriate concept and a broad stage for these four interactions. It claims to be able to study the physical properties of these four interactions simultaneously in the trigintaduonion spaces. Some scholars are steadily advancing towards the goal. This novel approach and cognition have expanded the scope of application of field theories.

(2) Composite fields. According to the viewpoint of trigintaduonion spaces, the weak nuclear field is essentially a composite field rather than a fundamental field. The weak nuclear field is composed of three fundamental fields, while the physical properties of these three fundamental fields are close to each other. According to the properties of trigintaduonions, the Yang-Mills equation, related to the weak nuclear fields, can be derived from these three fundamental fields of weak nuclear fields. This appears quite clear and smooth in the description. This method can provide a few deductions consistent with the conclusions of electroweak theory.

(3) Concise narrations. In the trigintaduonion spaces, the strong nuclear field is essentially one composite field rather than a fundamental field as well. The strong nuclear field consists of three fundamental fields, and the physical properties of these three fundamental fields are similar to each other. According to the properties of trigintaduonions, the Yang-Mills equation, relevant to the strong nuclear fields, can be given from these three fundamental fields of strong nuclear fields. This appears quite concise in logic. This method can achieve several inferences consistent with the conclusions of quark theory.

In the paper, it is capable of applying the algebra of trigintaduonions to explore the physical quantities of four interactions, including the trigintaduonion field potential, field strength, field source, linear momentum, angular momentum, torque, and force. The application of trigintaduonions is able to deduce a few inferences. The latter is consistent with the conclusions of the Yang-Mills equation of non-Abelian gauge fields, weak unified fields, and dark matter fields. The trigintaduonion field theory can solve some problems left over from the ``Standard Model" and even ``Beyond the Standard Model".

\section{\label{sec:level1}Fundamental fields}

J. C. Maxwell first utilized the algebra of quaternions to discuss the physical properties of electromagnetic fields. The method can be extended to other three fields, including the gravitational fields, weak nuclear fields, and strong nuclear fields. R. Descartes believed that the space is just the extension of substance. Nowadays, this point of view can be improved to that the fundamental space is only the extension of fundamental field. It means that each of fundamental fields possesses one fundamental space. And each fundamental space can be selected as a quaternion space.

Further, the gravitational field is distinct from the electromagnetic field, so that the quaternion space for gravitational fields is independent of that for electromagnetic fields. These two mutually independent quaternion spaces can constitute one octonion space. The octonion spaces are able to explore the physical properties of gravitational fields and electromagnetic fields simultaneously. This viewpoint can be extended to weak nuclear fields and strong nuclear fields.

By analyzing the electroweak theory, it can be concluded that the weak nuclear field is one type of composite field, which is made up of three fundamental fields independent of each other, including the weak nuclear field wa, weak nuclear field wb, weak nuclear field wc. Therefore, it is necessary to apply three independent quaternion spaces to describe the physical properties of weak nuclear fields. According to the Cayley-Dickson construction, the smallest multidimensional space (relevant to the quaternion spaces) that can accommodate the above five mutually independent quaternion spaces is the trigintaduonion space obviously. By means of the quark theory and multiplication table (Tables 1 and 2) of trigintaduonions, it is found that a strong nuclear field is a composite field, which consists of three mutually independent fundamental fields, including the strong nuclear field sa, strong nuclear field sb, strong nuclear field sc. Consequently, we have to utilize three independent and conjugate quaternion spaces to discuss the physical properties of strong nuclear fields.

These eight independent quaternion spaces constitute a trigintaduonion space. The application of trigintaduonion spaces is capable of exploring simultaneously the physical properties of electromagnetic fields, gravitational fields, weak nuclear fields, and strong nuclear fields.

\subsection{\label{sec:level1}Electromagnetic and gravitational fields}

In the electromagnetic and gravitational theories described by the octonions, the quaternion operator $\lozenge$ and the field strength of two fundamental fields (electromagnetic field and gravitational field) can be combined together to form a composite operator, $\lozenge ( \mathbb{F}_g , \mathbb{F}_e )$ . This has been discussed in some papers \cite{weng1,weng2}. Subsequently, this viewpoint can be further promoted. The quaternion operator can constitute a composite operator, $\lozenge ( \mathbb{F}_g , \mathbb{F}_e ; \mathbb{A}_g , \mathbb{A}_e)$ , together with the field potential and field strength of two fundamental fields (electromagnetic field and gravitational field). Herein $\mathbb{F}_g$ and $\mathbb{F}_e$ are the gravitational strength and electromagnetic strength, respectively. $\mathbb{A}_g$ and $\mathbb{A}_e$ are the gravitational potential and electromagnetic potential, respectively.

\subsection{\label{sec:level1}Weak nuclear fields}

From the point of view of the quaternion operator $\lozenge$ , the quaternion operator can form a composite operator, $\lozenge ( \mathbb{A}_e , \mathbb{A}_{wa} , \mathbb{A}_{wb} , \mathbb{A}_{wc} )$ , together with the field potentials of four fundamental fields (electromagnetic field, weak nuclear field wa, weak nuclear field wb, weak nuclear field wc), in the electroweak theory. Herein $\mathbb{A}_{wa}$ , $\mathbb{A}_{wb}$ , and $\mathbb{A}_{wc}$ are the field potentials of weak nuclear field wa, weak nuclear field wb, and weak nuclear field wc, respectively.

Three fundamental fields (weak nuclear field wa, weak nuclear field wb, weak nuclear field wc) collectively research the physical phenomena of weak interactions. Obviously, this method can be generalized. The quaternion operator $\lozenge$ can become a composite operator, $\lozenge ( \mathbb{F}_e , \mathbb{F}_{wa} , \mathbb{F}_{wb} , \mathbb{F}_{wc} ; \mathbb{A}_e , \mathbb{A}_{wa} , \mathbb{A}_{wb} , \mathbb{A}_{wc} )$ , together with the field potential and field strength of four fundamental fields, including the electromagnetic field, weak nuclear field wa, weak nuclear field wb, weak nuclear field wc. Herein $\mathbb{F}_{wa}$ , $\mathbb{F}_{wb}$ , and $\mathbb{F}_{wc}$ are the field strengths of weak nuclear field wa, weak nuclear field wb, and weak nuclear field wc, respectively.

It is easy to find that two composite operators, $\lozenge ( \mathbb{F}_g , \mathbb{F}_e ; \mathbb{A}_g , \mathbb{A}_e)$ and $\lozenge ( \mathbb{F}_e , \mathbb{F}_{wa} , \mathbb{F}_{wb} , \mathbb{F}_{wc} ; \mathbb{A}_e , \mathbb{A}_{wa} , \mathbb{A}_{wb} , \mathbb{A}_{wc} )$, both contain the quaternion operator $\lozenge$ and a fundamental field (electromagnetic field). Consequently, these two composite operators and the physical quantities of fundamental fields can be combined together to become one larger composite operator, $\lozenge ( \mathbb{F}_g , \mathbb{F}_e , \mathbb{F}_{wa} , \mathbb{F}_{wb} , \mathbb{F}_{wc} ; \mathbb{A}_g , \mathbb{A}_e , \mathbb{A}_{wa} , \mathbb{A}_{wb} , \mathbb{A}_{wc} )$. This larger composite operator consists of the quaternion operator and the physical quantities of five fundamental fields, including the gravitational field, electromagnetic field, weak nuclear field wa, weak nuclear field wb, weak nuclear field wc.

In the paper, the fundamental space is the extension of one fundamental field. Each fundamental field possesses one fundamental space, while each fundamental space is the quaternion space. As a result, these five fundamental fields possess five fundamental spaces. In other words, these five independent quaternion spaces can constitute a multidimensional space, which is relevant to the quaternion spaces. However, these five mutually independent quaternion spaces are unable to form a multidimensional space to satisfy the needs of multiplication table. According to the Cayley-Dickson construction and the multiplication tables relevant to these quaternions, the multiple quaternions independent of each other are able to constitute the octonions, sedenions, or trigintaduonions.

According to the viewpoint of trigintaduonions, the electroweak theory possesses three important characteristics. (1) Operators. The quaternion operator $\lozenge$ can be combined with the field potential $\mathbb{A}$ to constitute one composite operator. (2) Number of fundamental fields. It is necessary to increase the number of fundamental fields relevant to the weak interactions. A weak nuclear field may include a weak nuclear field wa, a weak nuclear field wb, and a weak nuclear field wc. It involves three fundamental fields and three fundamental spaces. (3) Symmetrical characteristics. The quaternions belong to the groups. However, the octonions belong to the rings rather than the groups. The sedenions and trigintaduonions are not the groups either.

The composite operator method in this article is essentially consistent with the viewpoint of electroweak theory in terms of operators and fundamental fields. This method can be extended to the strong nuclear fields.

\subsection{\label{sec:level1}Strong nuclear fields}

For these four interactions, the above composite operator, $\lozenge ( \mathbb{F}_g , \mathbb{F}_e , \mathbb{F}_{wa} , \mathbb{F}_{wb} , \mathbb{F}_{wc} ; \mathbb{A}_g , \mathbb{A}_e , \mathbb{A}_{wa} , \mathbb{A}_{wb} , \mathbb{A}_{wc} )$ , is unable to meet the requirements of the field theories. In order to satisfy the multiplication table of multidimensional spaces related to these quaternions, the composite operator must be expanded to comprise eight types of fundamental fields. This multidimensional space is a trigintaduonion space, and its composite operator is $\lozenge ( \mathbb{F}_g , \mathbb{F}_e , \mathbb{F}_{wa} , \mathbb{F}_{wb} , \mathbb{F}_{wc} , \mathbb{F}_{sa} , \mathbb{F}_{sb} , \mathbb{F}_{sc} ; \mathbb{A}_g , \mathbb{A}_e , \mathbb{A}_{wa} , \mathbb{A}_{wb} , \mathbb{A}_{wc} , \mathbb{A}_{sa} , \mathbb{A}_{sb} , \mathbb{A}_{sc} )$. Herein $\mathbb{A}_{sa}$ , $\mathbb{A}_{sb}$ , and $\mathbb{A}_{sc}$ are the field potentials of strong nuclear field sa, strong nuclear field sb, and strong nuclear field sc, respectively. $\mathbb{F}_{sa}$ , $\mathbb{F}_{sb}$ , and $\mathbb{F}_{sc}$ are the field strengths of strong nuclear field sa, strong nuclear field sb, and strong nuclear field sc, respectively.

In the trigintaduonion spaces, these eight types of fundamental fields consist of gravitational field, electromagnetic field, weak nuclear field wa, weak nuclear field wb, weak nuclear field wc, strong nuclear field sa, strong nuclear field sb, and strong nuclear field sc. Three fundamental fields (strong nuclear field sa, strong nuclear field sb, strong nuclear field sc) explore jointly the physical phenomena of strong interactions.

Further, in terms of the field theories relevant to the composite operators in the paper, the quaternion operator $\lozenge$ can be extended from the quaternion to the octonion, sedenion, and trigintaduonion, respectively.

\subsection{\label{sec:level1}Adjoint fields}

The fundamental fields involved in the electroweak theory are different from gravitational fields or strong nuclear fields. The electroweak theory requires expanding the number of fundamental fields to four. If the subsequent field theories are to contain completely the gravitational fields, the number of fundamental fields must be expanded from four to five. Considering the requirement of multiplication table of multidimensional spaces related to the quaternions, the paper adopts the research scheme of eight fundamental fields.

There are eight fundamental spaces independent of each other, according to the requirements of multiplication table of multidimensional spaces related to the quaternions. Each fundamental space is selected as a quaternion space. These eight independent fundamental spaces constitute one trigintaduonion space. The fundamental space is an extension of the fundamental field, so that there are eight fundamental fields independent of each other in the trigintaduonion spaces, including the gravitational field, electromagnetic field, weak nuclear field wa, weak nuclear field wb, weak nuclear field wc, strong nuclear field sa, strong nuclear field sb, and strong nuclear field sc.

In the trigintaduonion spaces, both the operator $\lozenge$ and the integrating function $\mathbb{X}$ of field potential $\mathbb{A}$ may be the trigintaduonion physical quantities. When the operator $\lozenge$ is one trigintaduonion physical quantity, according to the definition of field potential, there are thirty-two field potentials, including the field potential of eight fundamental fields, and the field potential of twenty-four adjoint fields.

If the trigintaduonion physical quantities are divided into each set of four dimensions, there are eight sets of four-dimensional physical quantities of fundamental fields. In other words, the adjoint field only causes one alteration in the spatial coordinate value of a fundamental field, and therefore does not vary radically the physical properties of the fundamental field. Each fundamental field possesses a few adjoint fields.

\subsection{\label{sec:level1}Relevant research}

The field equations in the octonion field theory, satisfied by the gravitational and electromagnetic fields, can be extended to these in the trigintaduonion field theory. However, the spatial dimensions involved in the strong and weak nuclear fields are too tiny, making it quite difficult to directly measure certain physical quantities at present. Therefore, it may be transformed into studying several relevant invariants, continuity equations, or equilibrium equations, for the strong and weak nuclear fields.

Considering the contribution of the curvature of trigintaduonion spaces to some physical quantities, it is able to achieve the trigintaduonion field theory in the curved spaces, including the Einstein's General Relativity. From the nondimensionalized trigintaduonion angular momentum, it is capable of inferring some inferences of quantum mechanics.

\begin{table}[b]
\caption{The multiplication table of trigintaduonions (part 1).}
\centering
\hbox{
\label{rotfloat3}%
\begin{tabular}[b]{c|cccc|cccc|cccc|cccc|}

\hline\hline

$ $ & $1$ & $\emph{\textbf{i}}_1$  & $\emph{\textbf{i}}_2$ & $\emph{\textbf{i}}_3$  & $\emph{\textbf{i}}_4$ & $\emph{\textbf{i}}_5$ & $\emph{\textbf{i}}_6$  & $\emph{\textbf{i}}_7$
& $\emph{\textbf{i}}_8$ & $\emph{\textbf{i}}_9$ & $\emph{\textbf{i}}_{10}$  & $\emph{\textbf{i}}_{11}$ & $\emph{\textbf{i}}_{12}$  & $\emph{\textbf{i}}_{13}$ & $\emph{\textbf{i}}_{14}$ & $\emph{\textbf{i}}_{15}$
%
%
\\
\hline
$1$ & $1$ & $\emph{\textbf{i}}_1$  & $\emph{\textbf{i}}_2$ & $\emph{\textbf{i}}_3$  & $\emph{\textbf{i}}_4$  & $\emph{\textbf{i}}_5$ & $\emph{\textbf{i}}_6$  & $\emph{\textbf{i}}_7$
& $\emph{\textbf{i}}_8$ & $\emph{\textbf{i}}_9$ & $\emph{\textbf{i}}_{10}$  & $\emph{\textbf{i}}_{11}$ & $\emph{\textbf{i}}_{12}$  & $\emph{\textbf{i}}_{13}$ & $\emph{\textbf{i}}_{14}$ & $\emph{\textbf{i}}_{15}$
%
%
\\
$\emph{\textbf{i}}_1$ & $\emph{\textbf{i}}_1$ & $-1$ & $\emph{\textbf{i}}_3$  & $-\emph{\textbf{i}}_2$ & $\emph{\textbf{i}}_5$ & $-\emph{\textbf{i}}_4$ &  $-\emph{\textbf{i}}_7$ & $\emph{\textbf{i}}_6$
& $\emph{\textbf{i}}_9$ & $-\emph{\textbf{i}}_8$ & $-\emph{\textbf{i}}_{11}$  & $\emph{\textbf{i}}_{10}$ & $-\emph{\textbf{i}}_{13}$  & $\emph{\textbf{i}}_{12}$ & $\emph{\textbf{i}}_{15}$ & $-\emph{\textbf{i}}_{14}$
%
%
\\
$\emph{\textbf{i}}_2$ & $\emph{\textbf{i}}_2$ & $-\emph{\textbf{i}}_3$ & $-1$ & $\emph{\textbf{i}}_1$  & $\emph{\textbf{i}}_6$  & $\emph{\textbf{i}}_7$ & $-\emph{\textbf{i}}_4$ & $-\emph{\textbf{i}}_5$
& $\emph{\textbf{i}}_{10}$ & $\emph{\textbf{i}}_{11}$ & $-\emph{\textbf{i}}_8$  & $-\emph{\textbf{i}}_9$ & $-\emph{\textbf{i}}_{14}$  & $-\emph{\textbf{i}}_{15}$ & $\emph{\textbf{i}}_{12}$ & $\emph{\textbf{i}}_{13}$
%
%
\\
$\emph{\textbf{i}}_3$ & $\emph{\textbf{i}}_3$ & $\emph{\textbf{i}}_2$ & $-\emph{\textbf{i}}_1$ & $-1$ & $\emph{\textbf{i}}_7$  & $-\emph{\textbf{i}}_6$ & $\emph{\textbf{i}}_5$  & $-\emph{\textbf{i}}_4$
& $\emph{\textbf{i}}_{11}$ & $-\emph{\textbf{i}}_{10}$ & $\emph{\textbf{i}}_9$  & $-\emph{\textbf{i}}_8$ & $-\emph{\textbf{i}}_{15}$  & $\emph{\textbf{i}}_{14}$ & $-\emph{\textbf{i}}_{13}$ & $\emph{\textbf{i}}_{12}$
%
%
\\
\hline
$\emph{\textbf{i}}_4$ & $\emph{\textbf{i}}_4$ & $-\emph{\textbf{i}}_5$ & $-\emph{\textbf{i}}_6$ & $-\emph{\textbf{i}}_7$ & $-1$ & $\emph{\textbf{i}}_1$ & $\emph{\textbf{i}}_2$  & $\emph{\textbf{i}}_3$
& $\emph{\textbf{i}}_{12}$ & $\emph{\textbf{i}}_{13}$ & $\emph{\textbf{i}}_{14}$  & $\emph{\textbf{i}}_{15}$ & $-\emph{\textbf{i}}_8$  & $-\emph{\textbf{i}}_9$ & $-\emph{\textbf{i}}_{10}$ & $-\emph{\textbf{i}}_{11}$
%
%
\\
$\emph{\textbf{i}}_5$ & $\emph{\textbf{i}}_5$ & $\emph{\textbf{i}}_4$ & $-\emph{\textbf{i}}_7$ & $\emph{\textbf{i}}_6$  & $-\emph{\textbf{i}}_1$ & $-1$ & $-\emph{\textbf{i}}_3$ & $\emph{\textbf{i}}_2$
& $\emph{\textbf{i}}_{13}$ & $-\emph{\textbf{i}}_{12}$ & $\emph{\textbf{i}}_{15}$  & $-\emph{\textbf{i}}_{14}$ & $\emph{\textbf{i}}_9$  & $-\emph{\textbf{i}}_8$ & $\emph{\textbf{i}}_{11}$ & $-\emph{\textbf{i}}_{10}$
%
%
\\
$\emph{\textbf{i}}_6$ & $\emph{\textbf{i}}_6$ & $\emph{\textbf{i}}_7$ & $\emph{\textbf{i}}_4$  & $-\emph{\textbf{i}}_5$ & $-\emph{\textbf{i}}_2$ & $\emph{\textbf{i}}_3$  & $-1$ & $-\emph{\textbf{i}}_1$
& $\emph{\textbf{i}}_{14}$ & $-\emph{\textbf{i}}_{15}$ & $-\emph{\textbf{i}}_{12}$  & $\emph{\textbf{i}}_{13}$ & $\emph{\textbf{i}}_{10}$  & $-\emph{\textbf{i}}_{11}$ & $-\emph{\textbf{i}}_8$ & $\emph{\textbf{i}}_9$
%
%
\\
$\emph{\textbf{i}}_7$ & $\emph{\textbf{i}}_7$ & $-\emph{\textbf{i}}_6$ & $\emph{\textbf{i}}_5$  & $\emph{\textbf{i}}_4$  & $-\emph{\textbf{i}}_3$ & $-\emph{\textbf{i}}_2$ & $\emph{\textbf{i}}_1$  & $-1$
& $\emph{\textbf{i}}_{15}$ & $\emph{\textbf{i}}_{14}$ & $-\emph{\textbf{i}}_{13}$  & $-\emph{\textbf{i}}_{12}$ & $\emph{\textbf{i}}_{11}$  & $\emph{\textbf{i}}_{10}$ & $-\emph{\textbf{i}}_9$ & $-\emph{\textbf{i}}_8$
%
%
\\
\hline
$\emph{\textbf{i}}_8$ & $\emph{\textbf{i}}_8$ & $-\emph{\textbf{i}}_9$ & $-\emph{\textbf{i}}_{10}$  & $-\emph{\textbf{i}}_{11}$  & $-\emph{\textbf{i}}_{12}$  & $-\emph{\textbf{i}}_{13}$ & $-\emph{\textbf{i}}_{14}$ & $-\emph{\textbf{i}}_{15}$
& $-1$ & $\emph{\textbf{i}}_1$  & $\emph{\textbf{i}}_2$ & $\emph{\textbf{i}}_3$ & $\emph{\textbf{i}}_4$  & $\emph{\textbf{i}}_5$ & $\emph{\textbf{i}}_6$  & $\emph{\textbf{i}}_7$
%
%
\\
$\emph{\textbf{i}}_9$ & $\emph{\textbf{i}}_9$ & $\emph{\textbf{i}}_8$ & $-\emph{\textbf{i}}_{11}$  & $\emph{\textbf{i}}_{10}$  & $-\emph{\textbf{i}}_{13}$  & $\emph{\textbf{i}}_{12}$ & $\emph{\textbf{i}}_{15}$ & $-\emph{\textbf{i}}_{14}$
& $-\emph{\textbf{i}}_1$ & $-1$ & $-\emph{\textbf{i}}_3$ & $\emph{\textbf{i}}_2$ & $-\emph{\textbf{i}}_5$  & $\emph{\textbf{i}}_4$ & $\emph{\textbf{i}}_7$  & $-\emph{\textbf{i}}_6$
%
%
\\
$\emph{\textbf{i}}_{10}$ & $\emph{\textbf{i}}_{10}$ & $\emph{\textbf{i}}_{11}$ & $\emph{\textbf{i}}_8$  & $-\emph{\textbf{i}}_9$  & $-\emph{\textbf{i}}_{14}$  & $-\emph{\textbf{i}}_{15}$ & $\emph{\textbf{i}}_{12}$ & $\emph{\textbf{i}}_{13}$
& $-\emph{\textbf{i}}_2$  & $\emph{\textbf{i}}_3$ & $-1$ & $-\emph{\textbf{i}}_1$ & $-\emph{\textbf{i}}_6$  & $-\emph{\textbf{i}}_7$ & $\emph{\textbf{i}}_4$  & $\emph{\textbf{i}}_5$
%
%
\\
$\emph{\textbf{i}}_{11}$ & $\emph{\textbf{i}}_{11}$ & $-\emph{\textbf{i}}_{10}$ & $\emph{\textbf{i}}_9$  & $\emph{\textbf{i}}_8$  & $-\emph{\textbf{i}}_{15}$  & $\emph{\textbf{i}}_{14}$ & $-\emph{\textbf{i}}_{13}$ & $\emph{\textbf{i}}_{12}$
& $-\emph{\textbf{i}}_3$  & $-\emph{\textbf{i}}_2$ & $\emph{\textbf{i}}_1$ & $-1$ & $-\emph{\textbf{i}}_7$  & $\emph{\textbf{i}}_6$ & $-\emph{\textbf{i}}_5$  & $\emph{\textbf{i}}_4$
%
%
\\
\hline
$\emph{\textbf{i}}_{12}$ & $\emph{\textbf{i}}_{12}$  & $\emph{\textbf{i}}_{13}$ & $\emph{\textbf{i}}_{14}$ & $\emph{\textbf{i}}_{15}$ & $\emph{\textbf{i}}_8$ & $-\emph{\textbf{i}}_9$ & $-\emph{\textbf{i}}_{10}$  & $-\emph{\textbf{i}}_{11}$
& $-\emph{\textbf{i}}_4$ & $\emph{\textbf{i}}_5$ & $\emph{\textbf{i}}_6$ & $\emph{\textbf{i}}_7$ & $-1$ & $-\emph{\textbf{i}}_1$ & $-\emph{\textbf{i}}_2$  & $-\emph{\textbf{i}}_3$
%
%
\\
$\emph{\textbf{i}}_{13}$ & $\emph{\textbf{i}}_{13}$  & $-\emph{\textbf{i}}_{12}$ & $\emph{\textbf{i}}_{15}$ & $-\emph{\textbf{i}}_{14}$ & $\emph{\textbf{i}}_9$ & $\emph{\textbf{i}}_8$ & $\emph{\textbf{i}}_{11}$  & $-\emph{\textbf{i}}_{10}$
& $-\emph{\textbf{i}}_5$ & $-\emph{\textbf{i}}_4$ & $\emph{\textbf{i}}_7$ & $-\emph{\textbf{i}}_6$ & $\emph{\textbf{i}}_1$ &$-1$ &  $\emph{\textbf{i}}_3$  & $-\emph{\textbf{i}}_2$
%
%
\\
$\emph{\textbf{i}}_{14}$ & $\emph{\textbf{i}}_{14}$  & $-\emph{\textbf{i}}_{15}$ & $-\emph{\textbf{i}}_{12}$ & $\emph{\textbf{i}}_{13}$ & $\emph{\textbf{i}}_{10}$ & $-\emph{\textbf{i}}_{11}$ & $\emph{\textbf{i}}_8$  & $\emph{\textbf{i}}_9$
& $-\emph{\textbf{i}}_6$ & $-\emph{\textbf{i}}_7$ & $-\emph{\textbf{i}}_4$ & $\emph{\textbf{i}}_5$ & $\emph{\textbf{i}}_2$ & $-\emph{\textbf{i}}_3$  & $-1$ & $\emph{\textbf{i}}_1$
%
%
\\
$\emph{\textbf{i}}_{15}$ & $\emph{\textbf{i}}_{15}$  & $\emph{\textbf{i}}_{14}$ & $-\emph{\textbf{i}}_{13}$ & $-\emph{\textbf{i}}_{12}$ & $\emph{\textbf{i}}_{11}$ & $\emph{\textbf{i}}_{10}$ & $-\emph{\textbf{i}}_9$  & $\emph{\textbf{i}}_8$
& $-\emph{\textbf{i}}_7$ & $\emph{\textbf{i}}_6$ & $-\emph{\textbf{i}}_5$ & $-\emph{\textbf{i}}_4$ & $\emph{\textbf{i}}_3$ & $\emph{\textbf{i}}_2$  & $-\emph{\textbf{i}}_1$ & $-1$
%
%

\\
\hline\hline
$\emph{\textbf{i}}_{16}$ & $\emph{\textbf{i}}_{16}$ & $-\emph{\textbf{i}}_{17}$  & $-\emph{\textbf{i}}_{18}$ & $-\emph{\textbf{i}}_{19}$  & $-\emph{\textbf{i}}_{20}$  & $-\emph{\textbf{i}}_{21}$ & $-\emph{\textbf{i}}_{22}$  & $-\emph{\textbf{i}}_{23}$
& $-\emph{\textbf{i}}_{24}$ & $-\emph{\textbf{i}}_{25}$ & $-\emph{\textbf{i}}_{26}$  & $-\emph{\textbf{i}}_{27}$ & $-\emph{\textbf{i}}_{28}$  & $-\emph{\textbf{i}}_{29}$ & $-\emph{\textbf{i}}_{30}$ & $-\emph{\textbf{i}}_{31}$
%
%
\\
$\emph{\textbf{i}}_{17}$ & $\emph{\textbf{i}}_{17}$ & $\emph{\textbf{i}}_{16}$ & $-\emph{\textbf{i}}_{19}$  & $\emph{\textbf{i}}_{18}$ & $-\emph{\textbf{i}}_{21}$ & $\emph{\textbf{i}}_{20}$ &  $\emph{\textbf{i}}_{23}$ & $-\emph{\textbf{i}}_{22}$
& $-\emph{\textbf{i}}_{25}$ & $\emph{\textbf{i}}_{24}$ & $\emph{\textbf{i}}_{27}$  & $-\emph{\textbf{i}}_{26}$ & $\emph{\textbf{i}}_{29}$  & $-\emph{\textbf{i}}_{28}$ & $-\emph{\textbf{i}}_{31}$ & $\emph{\textbf{i}}_{30}$
%
%
\\
$\emph{\textbf{i}}_{18}$ & $\emph{\textbf{i}}_{18}$ & $\emph{\textbf{i}}_{19}$ & $\emph{\textbf{i}}_{16}$ & $-\emph{\textbf{i}}_{17}$  & $-\emph{\textbf{i}}_{22}$  & $-\emph{\textbf{i}}_{23}$ & $\emph{\textbf{i}}_{20}$ & $\emph{\textbf{i}}_{21}$
& $-\emph{\textbf{i}}_{26}$ & $-\emph{\textbf{i}}_{27}$ & $\emph{\textbf{i}}_{24}$  & $\emph{\textbf{i}}_{25}$ & $\emph{\textbf{i}}_{30}$  & $\emph{\textbf{i}}_{31}$ & $-\emph{\textbf{i}}_{28}$ & $-\emph{\textbf{i}}_{29}$
%
%
\\
$\emph{\textbf{i}}_{19}$ & $\emph{\textbf{i}}_{19}$ & $-\emph{\textbf{i}}_{18}$ & $\emph{\textbf{i}}_{17}$ & $\emph{\textbf{i}}_{16}$ & $-\emph{\textbf{i}}_{23}$ & $\emph{\textbf{i}}_{22}$ & $-\emph{\textbf{i}}_{21}$  & $\emph{\textbf{i}}_{20}$
& $-\emph{\textbf{i}}_{27}$ & $\emph{\textbf{i}}_{26}$ & $-\emph{\textbf{i}}_{25}$  & $\emph{\textbf{i}}_{24}$ & $\emph{\textbf{i}}_{31}$  & $-\emph{\textbf{i}}_{30}$ & $\emph{\textbf{i}}_{29}$ & $-\emph{\textbf{i}}_{28}$
%
%
\\
\hline
$\emph{\textbf{i}}_{20}$ & $\emph{\textbf{i}}_{20}$ & $\emph{\textbf{i}}_{21}$ & $\emph{\textbf{i}}_{22}$ & $\emph{\textbf{i}}_{23}$ & $\emph{\textbf{i}}_{16}$ & $-\emph{\textbf{i}}_{17}$ & $-\emph{\textbf{i}}_{18}$  & $-\emph{\textbf{i}}_{19}$
& $-\emph{\textbf{i}}_{28}$ & $-\emph{\textbf{i}}_{29}$ & $-\emph{\textbf{i}}_{30}$  & $-\emph{\textbf{i}}_{31}$ & $\emph{\textbf{i}}_{24}$  & $\emph{\textbf{i}}_{25}$ & $\emph{\textbf{i}}_{26}$ & $\emph{\textbf{i}}_{27}$
%
%
\\
$\emph{\textbf{i}}_{21}$ & $\emph{\textbf{i}}_{21}$  & $-\emph{\textbf{i}}_{20}$ & $\emph{\textbf{i}}_{23}$  & $-\emph{\textbf{i}}_{22}$ & $\emph{\textbf{i}}_{17}$ & $\emph{\textbf{i}}_{16}$ & $\emph{\textbf{i}}_{19}$ & $-\emph{\textbf{i}}_{18}$
& $-\emph{\textbf{i}}_{29}$ & $\emph{\textbf{i}}_{28}$ & $-\emph{\textbf{i}}_{31}$  & $\emph{\textbf{i}}_{30}$ & $-\emph{\textbf{i}}_{25}$  & $\emph{\textbf{i}}_{24}$ & $-\emph{\textbf{i}}_{27}$ & $\emph{\textbf{i}}_{26}$
%
%
\\
$\emph{\textbf{i}}_{22}$ & $\emph{\textbf{i}}_{22}$  & $-\emph{\textbf{i}}_{23}$ & $-\emph{\textbf{i}}_{20}$  & $\emph{\textbf{i}}_{21}$ & $\emph{\textbf{i}}_{18}$ & $-\emph{\textbf{i}}_{19}$  & $\emph{\textbf{i}}_{16}$ & $\emph{\textbf{i}}_{17}$
& $-\emph{\textbf{i}}_{30}$ & $\emph{\textbf{i}}_{31}$ & $\emph{\textbf{i}}_{28}$  & $-\emph{\textbf{i}}_{29}$ & $-\emph{\textbf{i}}_{26}$  & $\emph{\textbf{i}}_{27}$ & $\emph{\textbf{i}}_{24}$ & $-\emph{\textbf{i}}_{25}$
%
%
\\
$\emph{\textbf{i}}_{23}$ & $\emph{\textbf{i}}_{23}$  & $\emph{\textbf{i}}_{22}$ & $-\emph{\textbf{i}}_{21}$  & $-\emph{\textbf{i}}_{20}$  & $\emph{\textbf{i}}_{19}$ & $\emph{\textbf{i}}_{18}$ & $-\emph{\textbf{i}}_{17}$  & $\emph{\textbf{i}}_{16}$
& $-\emph{\textbf{i}}_{31}$ & $-\emph{\textbf{i}}_{30}$ & $\emph{\textbf{i}}_{29}$  & $\emph{\textbf{i}}_{28}$ & $-\emph{\textbf{i}}_{27}$  & $-\emph{\textbf{i}}_{26}$ & $\emph{\textbf{i}}_{25}$ & $\emph{\textbf{i}}_{24}$
%
%
\\
\hline
$\emph{\textbf{i}}_{24}$ & $\emph{\textbf{i}}_{24}$  & $\emph{\textbf{i}}_{25}$ & $\emph{\textbf{i}}_{26}$  & $\emph{\textbf{i}}_{27}$  & $\emph{\textbf{i}}_{28}$  & $\emph{\textbf{i}}_{29}$ & $\emph{\textbf{i}}_{30}$ & $\emph{\textbf{i}}_{31}$
& $\emph{\textbf{i}}_{16}$ & $-\emph{\textbf{i}}_{17}$  & $-\emph{\textbf{i}}_{18}$ & $-\emph{\textbf{i}}_{19}$ & $-\emph{\textbf{i}}_{20}$  & $-\emph{\textbf{i}}_{21}$ & $-\emph{\textbf{i}}_{22}$  & $-\emph{\textbf{i}}_{23}$
%
%
\\
$\emph{\textbf{i}}_{25}$ & $\emph{\textbf{i}}_{25}$  & $-\emph{\textbf{i}}_{24}$ & $\emph{\textbf{i}}_{27}$  & $-\emph{\textbf{i}}_{26}$  & $\emph{\textbf{i}}_{29}$  & $-\emph{\textbf{i}}_{28}$ & $-\emph{\textbf{i}}_{31}$ & $\emph{\textbf{i}}_{30}$
& $\emph{\textbf{i}}_{17}$ & $\emph{\textbf{i}}_{16}$ & $\emph{\textbf{i}}_{19}$ & $-\emph{\textbf{i}}_{18}$ & $\emph{\textbf{i}}_{21}$  & $-\emph{\textbf{i}}_{20}$ & $-\emph{\textbf{i}}_{23}$  & $\emph{\textbf{i}}_{22}$
%
%
\\
$\emph{\textbf{i}}_{26}$ & $\emph{\textbf{i}}_{26}$  & $-\emph{\textbf{i}}_{27}$ & $-\emph{\textbf{i}}_{24}$  & $\emph{\textbf{i}}_{25}$  & $\emph{\textbf{i}}_{30}$  & $\emph{\textbf{i}}_{31}$ & $-\emph{\textbf{i}}_{28}$ & $-\emph{\textbf{i}}_{29}$
& $\emph{\textbf{i}}_{18}$  & $-\emph{\textbf{i}}_{19}$ & $\emph{\textbf{i}}_{16}$ & $\emph{\textbf{i}}_{17}$ & $\emph{\textbf{i}}_{22}$  & $\emph{\textbf{i}}_{23}$ & $-\emph{\textbf{i}}_{20}$  & $-\emph{\textbf{i}}_{21}$
%
%
\\
$\emph{\textbf{i}}_{27}$ & $\emph{\textbf{i}}_{27}$  & $\emph{\textbf{i}}_{26}$ & $-\emph{\textbf{i}}_{25}$  & $-\emph{\textbf{i}}_{24}$  & $\emph{\textbf{i}}_{31}$  & $-\emph{\textbf{i}}_{30}$ & $\emph{\textbf{i}}_{29}$ & $-\emph{\textbf{i}}_{28}$
& $\emph{\textbf{i}}_{19}$  & $\emph{\textbf{i}}_{18}$ & $-\emph{\textbf{i}}_{17}$ & $\emph{\textbf{i}}_{16}$ & $\emph{\textbf{i}}_{23}$  & $-\emph{\textbf{i}}_{22}$ & $\emph{\textbf{i}}_{21}$  & $-\emph{\textbf{i}}_{20}$
%
%
\\
\hline
$\emph{\textbf{i}}_{28}$ & $\emph{\textbf{i}}_{28}$   & $-\emph{\textbf{i}}_{29}$ & $-\emph{\textbf{i}}_{30}$ & $-\emph{\textbf{i}}_{31}$ & $-\emph{\textbf{i}}_{24}$ & $\emph{\textbf{i}}_{25}$ & $\emph{\textbf{i}}_{26}$  & $\emph{\textbf{i}}_{27}$
& $\emph{\textbf{i}}_{20}$ & $-\emph{\textbf{i}}_{21}$ & $-\emph{\textbf{i}}_{22}$ & $-\emph{\textbf{i}}_{23}$ & $\emph{\textbf{i}}_{16}$ & $\emph{\textbf{i}}_{17}$ & $\emph{\textbf{i}}_{18}$  & $\emph{\textbf{i}}_{19}$
%
%
\\
$\emph{\textbf{i}}_{29}$ & $\emph{\textbf{i}}_{29}$   & $\emph{\textbf{i}}_{28}$ & $-\emph{\textbf{i}}_{31}$ & $\emph{\textbf{i}}_{30}$ & $-\emph{\textbf{i}}_{25}$ & $-\emph{\textbf{i}}_{24}$ & $-\emph{\textbf{i}}_{27}$  & $\emph{\textbf{i}}_{26}$
& $\emph{\textbf{i}}_{21}$ & $\emph{\textbf{i}}_{20}$ & $-\emph{\textbf{i}}_{23}$ & $\emph{\textbf{i}}_{22}$ & $-\emph{\textbf{i}}_{17}$ & $\emph{\textbf{i}}_{16}$ &  $-\emph{\textbf{i}}_{19}$  & $\emph{\textbf{i}}_{18}$
%
%
\\
$\emph{\textbf{i}}_{30}$ & $\emph{\textbf{i}}_{30}$   & $\emph{\textbf{i}}_{31}$ & $\emph{\textbf{i}}_{28}$ & $-\emph{\textbf{i}}_{29}$ & $-\emph{\textbf{i}}_{26}$ & $\emph{\textbf{i}}_{27}$ & $-\emph{\textbf{i}}_{24}$  & $-\emph{\textbf{i}}_{25}$
& $\emph{\textbf{i}}_{22}$ & $\emph{\textbf{i}}_{23}$ & $\emph{\textbf{i}}_{20}$ & $-\emph{\textbf{i}}_{21}$ & $-\emph{\textbf{i}}_{18}$ & $\emph{\textbf{i}}_{19}$  & $\emph{\textbf{i}}_{16}$ & $-\emph{\textbf{i}}_{17}$
%
%
\\
$\emph{\textbf{i}}_{31}$ & $\emph{\textbf{i}}_{31}$  & $-\emph{\textbf{i}}_{30}$ & $\emph{\textbf{i}}_{29}$ & $\emph{\textbf{i}}_{28}$ & $-\emph{\textbf{i}}_{27}$ & $-\emph{\textbf{i}}_{26}$ & $\emph{\textbf{i}}_{25}$  & $-\emph{\textbf{i}}_{24}$
& $\emph{\textbf{i}}_{23}$ & $-\emph{\textbf{i}}_{22}$ & $\emph{\textbf{i}}_{21}$ & $\emph{\textbf{i}}_{20}$ & $-\emph{\textbf{i}}_{19}$ & $-\emph{\textbf{i}}_{18}$  & $\emph{\textbf{i}}_{17}$ & $\emph{\textbf{i}}_{16}$
%
%
\\
\hline\hline
\end{tabular}
}
\end{table}

\section{\label{sec:level1}Trigintaduonion spaces}

The trigintaduonion spaces can be decomposed into some subspaces independent of each other, including eight mutually independent four-dimensional spaces, $\mathbb{H}_g$ , $\mathbb{H}_e$ , $\mathbb{H}_{wa}$ , $\mathbb{H}_{wb}$ , $\mathbb{H}_{wc}$ , $\mathbb{H}_{sa}$ , $\mathbb{H}_{sb}$ , $\mathbb{H}_{sc}$ .

The first subspace, $\mathbb{H}_g$ , can be utilized to explore the physical properties of gravitational fields. And it is one quaternion space. In the quaternion space, $\mathbb{H}_g$ , for gravitational fields, the coordinate values are $R_{g0}$ and $R_{gk}$, and the basis vector is $\textbf{I}_{gj}$ . The four-dimensional radius vector is, $\mathbb{R}_g = \textrm{i} \textbf{I}_{g0} R_{g0} + \Sigma \textbf{I}_{gk} R_{gk}$ . Herein $R_{gj}$ is real. $R_{g0} = v_0 t$. $v_0$ is the speed of light, while $t$ is the time. $\textbf{I}_{g0} = 1$. $\textbf{I}_{g0} \circ \textbf{I}_{g0} = 1$. $\textbf{I}_{gk} \circ \textbf{I}_{gk} = -1$. $\textrm{i}$ is the imaginary unit. $j = 0, 1, 2, 3$. $k = 1, 2, 3$.

The second subspace, $\mathbb{H}_e$ , is able to be applied to research the physical properties of electromagnetic fields. In essence, it is also a quaternion space. In the subspace, $\mathbb{H}_e$ , for electromagnetic fields, the coordinate values are $R_{e0}$ and $R_{ek}$ , and the basis vector is $\textbf{I}_{ej}$ . The four-dimensional radius vector is, $\mathbb{R}_e = \textrm{i} \textbf{I}_{e0} R_{e0} + \Sigma \textbf{I}_{ek} R_{ek}$. Two subspaces, $\mathbb{H}_g$ and $\mathbb{H}_e$ , can be included in one octonion space $\mathbb{O}$ . Herein $R_{ej}$ is real. $\textbf{I}_{ej} = \textbf{I}_{gj} \circ \textbf{I}_{e0}$ . $\textbf{I}_{ej} \circ \textbf{I}_{ej} = -1$.

The physical properties of weak nuclear fields are relatively complicated, and they require the use of three four-dimensional subspaces to be clearly described. Three four-dimensional subspaces, $\mathbb{H}_{wa}$ , $\mathbb{H}_{wb}$ , $\mathbb{H}_{wc}$ , can be utilized to explore the physical properties of weak nuclear fields. These three subspaces are the quaternion spaces essentially. The application of three four-dimensional subspaces can effectively study the physical properties of weak nuclear fields, and their inferences are consistent with the conclusions of electroweak theory.

(1) In the subspace space, $\mathbb{H}_{wa}$ , for the weak nuclear field wa , the coordinate values are $R_{wa0}$ and $R_{wak}$ , and the basis vector is $\textbf{I}_{waj}$ . The four-dimensional radius vector is, $\mathbb{R}_{wa} = \textrm{i} \textbf{I}_{wa0} R_{wa0} + \Sigma \textbf{I}_{wak} R_{wak}$ . Herein $R_{waj}$ is real. $\textbf{I}_{waj} = \textbf{I}_{gj} \circ \textbf{I}_{wa0}$ . $\textbf{I}_{waj} \circ \textbf{I}_{waj} = - 1$.

(2) In the subspace space, $\mathbb{H}_{wb}$ , for the weak nuclear field wb , the coordinate values are $R_{wb0}$ and $R_{wbk}$ , and the basis vector is $\textbf{I}_{wbj}$ . The four-dimensional radius vector is, $\mathbb{R}_{wb} = \textrm{i} \textbf{I}_{wb0} R_{wb0} + \Sigma \textbf{I}_{wbk} R_{wbk}$ . Herein $R_{wbj}$ is real. $\textbf{I}_{wbj} = \textbf{I}_{gj} \circ \textbf{I}_{wb0}$ . $\textbf{I}_{wbj} \circ \textbf{I}_{wbj} = - 1$.

(3) In the subspace space, $\mathbb{H}_{wc}$ , for the weak nuclear field wc , the coordinate values are $R_{wc0}$ and $R_{wck}$ , and the basis vector is $\textbf{I}_{wcj}$ . The four-dimensional radius vector is, $\mathbb{R}_{wc} = \textrm{i} \textbf{I}_{wc0} R_{wc0} + \Sigma \textbf{I}_{wck} R_{wck}$ . Herein $R_{wcj}$ is real. $\textbf{I}_{wcj} = \textbf{I}_{gj} \circ \textbf{I}_{wc0}$ . $\textbf{I}_{wcj} \circ \textbf{I}_{wcj} = - 1$.

The physical properties of strong nuclear fields are relatively complicated, and it requires the application of three four-dimensional subspaces to describe them clearly. Three four-dimensional subspaces, $\mathbb{H}_{sa}$ , $\mathbb{H}_{sb}$ , $\mathbb{H}_{sc}$ , can be utilized to explore the physical properties of strong nuclear fields. These three subspaces are the conjugate quaternion spaces essentially. The application of three four-dimensional subspaces can effectively investigate the physical properties of strong nuclear fields, and the obtained inferences are consistent with the conclusions of quark theory.

(1) In the subspace space, $\mathbb{H}_{sa}$ , for the strong nuclear field sa , the coordinate values are $R_{sa0}$ and $R_{sak}$ , and the basis vector is $\textbf{I}_{saj}$ . The four-dimensional radius vector can be written as, $\mathbb{R}_{sa} = \textrm{i} \textbf{I}_{sa0} R_{sa0} + \Sigma \textbf{I}_{sak} R_{sak}$ . Herein $R_{saj}$ is real. $\textbf{I}_{saj} = \textbf{I}_{gj}^\ast \circ \textbf{I}_{sa0}$ . $\textbf{I}_{saj} \circ \textbf{I}_{saj} = - 1$.

(2) In the subspace space, $\mathbb{H}_{sb}$ , for the strong nuclear field sb , the coordinate values are $R_{sb0}$ and $R_{sbk}$ , and the basis vector is $\textbf{I}_{sbj}$ . The four-dimensional radius vector can be written as, $\mathbb{R}_{sb} = \textrm{i} \textbf{I}_{sb0} R_{sb0} + \Sigma \textbf{I}_{sbk} R_{sbk}$ . Herein $R_{sbj}$ is real. $\textbf{I}_{sbj} = \textbf{I}_{gj}^\ast \circ \textbf{I}_{sb0}$ . $\textbf{I}_{sbj} \circ \textbf{I}_{sbj} = - 1$.

(3) In the subspace space, $\mathbb{H}_{sc}$ , for the strong nuclear field sc , the coordinate values are $R_{sc0}$ and $R_{sck}$ , and the basis vector is $\textbf{I}_{scj}$ . The four-dimensional radius vector can be written as, $\mathbb{R}_{sc} = \textrm{i} \textbf{I}_{sc0} R_{sc0} + \Sigma \textbf{I}_{sck} R_{sck}$ . Herein $R_{scj}$ is real. $\textbf{I}_{scj} = \textbf{I}_{gj}^\ast \circ \textbf{I}_{sc0}$ . $\textbf{I}_{scj} \circ \textbf{I}_{scj} = - 1$.

In the trigintaduonion spaces, these eight radius vectors in the above can be combined into one trigintaduonion radius vector,
\begin{eqnarray}
\mathbb{R} = \mathbb{R}_g + k_{eg} \mathbb{R}_e + k_{wag} \mathbb{R}_{wa} + k_{wbg} \mathbb{R}_{wb} + k_{wcg} \mathbb{R}_{wc} + k_{sag} \mathbb{R}_{sa} + k_{sbg} \mathbb{R}_{sb} + k_{scg} \mathbb{R}_{sc} ~ ,
\end{eqnarray}
where $k_{eg}$ , $k_{wag}$ , $k_{wbg}$ , $k_{wcg}$ , $k_{sag}$ , $k_{sbg}$ , and $k_{scg}$ are coefficients, to meet the requirement of dimensional homogeneity.

In order to more concisely express the trigintaduonion radius vector, we can also utilize one new marking method to narrate the above basis vectors. (a) $\textbf{I}_{gj} = \textbf{I}_j $ . (b) $\textbf{I}_{ej} = \textbf{I}_{j+4} $ . (c) $\textbf{I}_{waj} = \textbf{I}_{j+8} $ . $\textbf{I}_{wbj} = \textbf{I}_{j+16} $ . $\textbf{I}_{wcj} = \textbf{I}_{j+28} $. (d) $\textbf{I}_{saj} = \textbf{I}_{j+12} $. $\textbf{I}_{sbj} = \textbf{I}_{j+20} $. $\textbf{I}_{scj} = \textbf{I}_{j+24} $ . Meanwhile the new labeling methods can be used to label these coordinate values above. (a) $R_{gj} = R_j$. (b) $R_{ej} = R_{j+4}$ . (c) $R_{waj} = R_{j+8}$ . $R_{wbj} = R_{j+16}$ . $R_{wcj} = R_{j+28}$ . (d) $R_{saj} = R_{j+12}$ . $R_{sbj} = R_{j+20}$ . $R_{scj} = R_{j+24}$ . As a result, the trigintaduonion radius vector can also be rewritten in a more concise way.

Similarly, it is capable of defining the trigintaduonion velocity $\mathbb{V}$ , field potential $\mathbb{A}$ , field strength $\mathbb{F}$ , and field source $\mathbb{S}$ and others in the trigintaduonion spaces.

In summary, it is necessary to introduce eight four-dimensional spaces independent of each other, in order to describe the physical properties of four interactions. They are closely related to the quaternion spaces. (a) The first quaternion space, $\textbf{I}_j$ , is utilized to describe the physical properties of gravitational fields. (b) The second quaternion space, $\textbf{I}_{j+4}$ , is used to express the physical properties of electromagnetic fields. (c) Three mutually independent four-dimensional spaces, $\textbf{I}_{j+8}$ , $\textbf{I}_{j+16}$ , $\textbf{I}_{j+28}$ , are applied to discuss the physical properties of weak nuclear fields. (d) Three mutually independent four-dimensional spaces, $\textbf{I}_{j+12}$, $\textbf{I}_{j+20}$, $\textbf{I}_{j+24}$ , are applied to depict the physical properties of strong nuclear fields. These eight independent four-dimensional spaces can be included in one trigintaduonion space, which meets the multiplication table of trigintaduonion spaces.

Obviously, one can observe the trigintaduonion space from other perspectives. In the complex spaces, the coordinate value corresponding to each basis vector can contain two real numbers. This viewpoint has been extended to the octonion spaces \cite{weng3}. According to the octonion multiplication table, an octonion can be considered as having four basis vectors. The coordinate value corresponding to each basis vector comprises two real numbers and even complex numbers. Further, this point of view can be extended to the sedenion spaces and trigintaduonion spaces. According to the trigintaduonion multiplication table, one trigintaduonion can be considered as possessing four basis vectors too. And the coordinate value corresponding to each basis vector incorporates eight real numbers and even complex numbers.

\begin{table}[b]
\caption{The multiplication table of trigintaduonions (part 2).}
\centering
\hbox{
\label{rotfloat3}%
\begin{tabular}[b]{c|cccc|cccc|cccc|ccccc}

\hline\hline

$ $ %
& $\emph{\textbf{i}}_{16}$ & $\emph{\textbf{i}}_{17}$  & $\emph{\textbf{i}}_{18}$ & $\emph{\textbf{i}}_{19}$  & $\emph{\textbf{i}}_{20}$ & $\emph{\textbf{i}}_{21}$ & $\emph{\textbf{i}}_{22}$  & $\emph{\textbf{i}}_{23}$
& $\emph{\textbf{i}}_{24}$ & $\emph{\textbf{i}}_{25}$ & $\emph{\textbf{i}}_{26}$  & $\emph{\textbf{i}}_{27}$ & $\emph{\textbf{i}}_{28}$  & $\emph{\textbf{i}}_{29}$ & $\emph{\textbf{i}}_{30}$ & $\emph{\textbf{i}}_{31}$
\\
\hline
$1$ %
& $\emph{\textbf{i}}_{16}$ & $\emph{\textbf{i}}_{17}$  & $\emph{\textbf{i}}_{18}$ & $\emph{\textbf{i}}_{19}$  & $\emph{\textbf{i}}_{20}$ & $\emph{\textbf{i}}_{21}$ & $\emph{\textbf{i}}_{22}$  & $\emph{\textbf{i}}_{23}$
& $\emph{\textbf{i}}_{24}$ & $\emph{\textbf{i}}_{25}$ & $\emph{\textbf{i}}_{26}$  & $\emph{\textbf{i}}_{27}$ & $\emph{\textbf{i}}_{28}$  & $\emph{\textbf{i}}_{29}$ & $\emph{\textbf{i}}_{30}$ & $\emph{\textbf{i}}_{31}$
\\
$\emph{\textbf{i}}_1$ %
& $\emph{\textbf{i}}_{17}$ & $-\emph{\textbf{i}}_{16}$  & $-\emph{\textbf{i}}_{19}$ & $\emph{\textbf{i}}_{18}$  & $-\emph{\textbf{i}}_{21}$ & $\emph{\textbf{i}}_{20}$ & $\emph{\textbf{i}}_{23}$  & $-\emph{\textbf{i}}_{22}$
& $-\emph{\textbf{i}}_{25}$ & $\emph{\textbf{i}}_{24}$ & $\emph{\textbf{i}}_{27}$  & $-\emph{\textbf{i}}_{26}$ & $\emph{\textbf{i}}_{29}$  & $-\emph{\textbf{i}}_{28}$ & $-\emph{\textbf{i}}_{31}$ & $\emph{\textbf{i}}_{30}$
\\
$\emph{\textbf{i}}_2$ %
& $\emph{\textbf{i}}_{18}$ & $\emph{\textbf{i}}_{19}$  & $-\emph{\textbf{i}}_{16}$ & $-\emph{\textbf{i}}_{17}$  & $-\emph{\textbf{i}}_{22}$ & $-\emph{\textbf{i}}_{23}$ & $\emph{\textbf{i}}_{20}$  & $\emph{\textbf{i}}_{21}$
& $-\emph{\textbf{i}}_{26}$ & $-\emph{\textbf{i}}_{27}$ & $\emph{\textbf{i}}_{24}$  & $\emph{\textbf{i}}_{25}$ & $\emph{\textbf{i}}_{30}$  & $\emph{\textbf{i}}_{31}$ & $-\emph{\textbf{i}}_{28}$ & $-\emph{\textbf{i}}_{29}$
\\
$\emph{\textbf{i}}_3$ %
& $\emph{\textbf{i}}_{19}$ & $-\emph{\textbf{i}}_{18}$  & $\emph{\textbf{i}}_{17}$ & $-\emph{\textbf{i}}_{16}$  & $-\emph{\textbf{i}}_{23}$ & $\emph{\textbf{i}}_{22}$ & $-\emph{\textbf{i}}_{21}$  & $\emph{\textbf{i}}_{20}$
& $-\emph{\textbf{i}}_{27}$ & $\emph{\textbf{i}}_{26}$ & $-\emph{\textbf{i}}_{25}$  & $\emph{\textbf{i}}_{24}$ & $\emph{\textbf{i}}_{31}$  & $-\emph{\textbf{i}}_{30}$ & $\emph{\textbf{i}}_{29}$ & $-\emph{\textbf{i}}_{28}$
\\
\hline
$\emph{\textbf{i}}_4$ %
& $\emph{\textbf{i}}_{20}$ & $\emph{\textbf{i}}_{21}$  & $\emph{\textbf{i}}_{22}$ & $\emph{\textbf{i}}_{23}$  & $-\emph{\textbf{i}}_{16}$ & $-\emph{\textbf{i}}_{17}$ & $-\emph{\textbf{i}}_{18}$  & $-\emph{\textbf{i}}_{19}$
& $-\emph{\textbf{i}}_{28}$ & $-\emph{\textbf{i}}_{29}$ & $-\emph{\textbf{i}}_{30}$  & $-\emph{\textbf{i}}_{31}$ & $\emph{\textbf{i}}_{24}$  & $\emph{\textbf{i}}_{25}$ & $\emph{\textbf{i}}_{26}$ & $\emph{\textbf{i}}_{27}$
\\
$\emph{\textbf{i}}_5$ %
& $\emph{\textbf{i}}_{21}$ & $-\emph{\textbf{i}}_{20}$  & $\emph{\textbf{i}}_{23}$ & $-\emph{\textbf{i}}_{22}$  & $\emph{\textbf{i}}_{17}$ & $-\emph{\textbf{i}}_{16}$ & $\emph{\textbf{i}}_{19}$  & $-\emph{\textbf{i}}_{18}$
& $-\emph{\textbf{i}}_{29}$ & $\emph{\textbf{i}}_{28}$ & $-\emph{\textbf{i}}_{31}$  & $\emph{\textbf{i}}_{30}$ & $-\emph{\textbf{i}}_{25}$  & $\emph{\textbf{i}}_{24}$ & $-\emph{\textbf{i}}_{27}$ & $\emph{\textbf{i}}_{26}$
\\
$\emph{\textbf{i}}_6$ %
& $\emph{\textbf{i}}_{22}$ & $-\emph{\textbf{i}}_{23}$  & $-\emph{\textbf{i}}_{20}$ & $\emph{\textbf{i}}_{21}$  & $\emph{\textbf{i}}_{18}$ & $-\emph{\textbf{i}}_{19}$ & $-\emph{\textbf{i}}_{16}$  & $\emph{\textbf{i}}_{17}$
& $-\emph{\textbf{i}}_{30}$ & $\emph{\textbf{i}}_{31}$ & $\emph{\textbf{i}}_{28}$  & $-\emph{\textbf{i}}_{29}$ & $-\emph{\textbf{i}}_{26}$  & $\emph{\textbf{i}}_{27}$ & $\emph{\textbf{i}}_{24}$ & $-\emph{\textbf{i}}_{25}$
\\
$\emph{\textbf{i}}_7$ %
& $\emph{\textbf{i}}_{23}$ & $\emph{\textbf{i}}_{22}$  & $-\emph{\textbf{i}}_{21}$ & $-\emph{\textbf{i}}_{20}$  & $\emph{\textbf{i}}_{19}$ & $\emph{\textbf{i}}_{18}$ & $-\emph{\textbf{i}}_{17}$  & $-\emph{\textbf{i}}_{16}$
& $-\emph{\textbf{i}}_{31}$ & $-\emph{\textbf{i}}_{30}$ & $\emph{\textbf{i}}_{29}$  & $\emph{\textbf{i}}_{28}$ & $-\emph{\textbf{i}}_{27}$  & $-\emph{\textbf{i}}_{26}$ & $\emph{\textbf{i}}_{25}$ & $\emph{\textbf{i}}_{24}$
\\
\hline
$\emph{\textbf{i}}_8$ %
& $\emph{\textbf{i}}_{24}$ & $\emph{\textbf{i}}_{25}$ & $\emph{\textbf{i}}_{26}$  & $\emph{\textbf{i}}_{27}$ & $\emph{\textbf{i}}_{28}$  & $\emph{\textbf{i}}_{29}$ & $\emph{\textbf{i}}_{30}$ & $\emph{\textbf{i}}_{31}$
& $-\emph{\textbf{i}}_{16}$ & $-\emph{\textbf{i}}_{17}$ & $-\emph{\textbf{i}}_{18}$  & $-\emph{\textbf{i}}_{19}$ & $-\emph{\textbf{i}}_{20}$  & $-\emph{\textbf{i}}_{21}$ & $-\emph{\textbf{i}}_{22}$ & $-\emph{\textbf{i}}_{23}$
\\
$\emph{\textbf{i}}_9$ %
& $\emph{\textbf{i}}_{25}$ & $-\emph{\textbf{i}}_{24}$  & $\emph{\textbf{i}}_{27}$ & $-\emph{\textbf{i}}_{26}$  & $\emph{\textbf{i}}_{29}$ & $-\emph{\textbf{i}}_{28}$ & $-\emph{\textbf{i}}_{31}$  & $\emph{\textbf{i}}_{30}$
& $\emph{\textbf{i}}_{17}$ & $-\emph{\textbf{i}}_{16}$ & $\emph{\textbf{i}}_{19}$  & $-\emph{\textbf{i}}_{18}$ & $\emph{\textbf{i}}_{21}$  & $-\emph{\textbf{i}}_{20}$ & $-\emph{\textbf{i}}_{23}$ & $\emph{\textbf{i}}_{22}$
\\
$\emph{\textbf{i}}_{10}$ %
& $\emph{\textbf{i}}_{26}$ & $-\emph{\textbf{i}}_{27}$  & $-\emph{\textbf{i}}_{24}$ & $\emph{\textbf{i}}_{25}$  & $\emph{\textbf{i}}_{30}$ & $\emph{\textbf{i}}_{31}$ & $-\emph{\textbf{i}}_{28}$  & $-\emph{\textbf{i}}_{29}$
& $\emph{\textbf{i}}_{18}$ & $-\emph{\textbf{i}}_{19}$ & $-\emph{\textbf{i}}_{16}$  & $\emph{\textbf{i}}_{17}$ & $\emph{\textbf{i}}_{22}$  & $\emph{\textbf{i}}_{23}$ & $-\emph{\textbf{i}}_{20}$ & $-\emph{\textbf{i}}_{21}$
\\
$\emph{\textbf{i}}_{11}$ %
& $\emph{\textbf{i}}_{27}$ & $\emph{\textbf{i}}_{26}$  & $-\emph{\textbf{i}}_{25}$ & $-\emph{\textbf{i}}_{24}$  & $\emph{\textbf{i}}_{31}$ & $-\emph{\textbf{i}}_{30}$ & $\emph{\textbf{i}}_{29}$  & $-\emph{\textbf{i}}_{28}$
& $\emph{\textbf{i}}_{19}$ & $\emph{\textbf{i}}_{18}$ & $-\emph{\textbf{i}}_{17}$  & $-\emph{\textbf{i}}_{16}$ & $\emph{\textbf{i}}_{23}$  & $-\emph{\textbf{i}}_{22}$ & $\emph{\textbf{i}}_{21}$ & $-\emph{\textbf{i}}_{20}$
\\
\hline
$\emph{\textbf{i}}_{12}$ %
& $\emph{\textbf{i}}_{28}$  & $-\emph{\textbf{i}}_{29}$ & $-\emph{\textbf{i}}_{30}$ & $-\emph{\textbf{i}}_{31}$  & $-\emph{\textbf{i}}_{24}$ & $\emph{\textbf{i}}_{25}$ & $\emph{\textbf{i}}_{26}$  & $\emph{\textbf{i}}_{27}$
& $\emph{\textbf{i}}_{20}$ & $-\emph{\textbf{i}}_{21}$ & $-\emph{\textbf{i}}_{22}$  & $-\emph{\textbf{i}}_{23}$ & $-\emph{\textbf{i}}_{16}$  & $\emph{\textbf{i}}_{17}$ & $\emph{\textbf{i}}_{18}$ & $\emph{\textbf{i}}_{19}$
\\
$\emph{\textbf{i}}_{13}$ %
& $\emph{\textbf{i}}_{29}$ & $\emph{\textbf{i}}_{28}$  & $-\emph{\textbf{i}}_{31}$ & $\emph{\textbf{i}}_{30}$  & $-\emph{\textbf{i}}_{25}$ & $-\emph{\textbf{i}}_{24}$ & $-\emph{\textbf{i}}_{27}$  & $\emph{\textbf{i}}_{26}$
& $\emph{\textbf{i}}_{21}$ & $\emph{\textbf{i}}_{20}$ & $-\emph{\textbf{i}}_{23}$  & $\emph{\textbf{i}}_{22}$ & $-\emph{\textbf{i}}_{17}$  & $-\emph{\textbf{i}}_{16}$ & $-\emph{\textbf{i}}_{19}$ & $\emph{\textbf{i}}_{18}$
\\
$\emph{\textbf{i}}_{14}$ %
& $\emph{\textbf{i}}_{30}$ & $\emph{\textbf{i}}_{31}$  & $\emph{\textbf{i}}_{28}$ & $-\emph{\textbf{i}}_{29}$  & $-\emph{\textbf{i}}_{26}$ & $\emph{\textbf{i}}_{27}$ & $-\emph{\textbf{i}}_{24}$  & $-\emph{\textbf{i}}_{25}$
& $\emph{\textbf{i}}_{22}$ & $\emph{\textbf{i}}_{23}$ & $\emph{\textbf{i}}_{20}$  & $-\emph{\textbf{i}}_{21}$ & $-\emph{\textbf{i}}_{18}$  & $\emph{\textbf{i}}_{19}$ & $-\emph{\textbf{i}}_{16}$ & $-\emph{\textbf{i}}_{17}$
\\
$\emph{\textbf{i}}_{15}$ %
& $\emph{\textbf{i}}_{31}$ & $-\emph{\textbf{i}}_{30}$  & $\emph{\textbf{i}}_{29}$ & $\emph{\textbf{i}}_{28}$  & $-\emph{\textbf{i}}_{27}$ & $-\emph{\textbf{i}}_{26}$ & $\emph{\textbf{i}}_{25}$  & $-\emph{\textbf{i}}_{24}$
& $\emph{\textbf{i}}_{23}$ & $-\emph{\textbf{i}}_{22}$ & $\emph{\textbf{i}}_{21}$  & $\emph{\textbf{i}}_{20}$ & $-\emph{\textbf{i}}_{19}$  & $-\emph{\textbf{i}}_{18}$ & $\emph{\textbf{i}}_{17}$ & $-\emph{\textbf{i}}_{16}$

\\
\hline\hline
$\emph{\textbf{i}}_{16}$ %
& $-1$ & $\emph{\textbf{i}}_1$  & $\emph{\textbf{i}}_2$ & $\emph{\textbf{i}}_3$  & $\emph{\textbf{i}}_4$ & $\emph{\textbf{i}}_5$ & $\emph{\textbf{i}}_6$  & $\emph{\textbf{i}}_7$
& $\emph{\textbf{i}}_8$ & $\emph{\textbf{i}}_9$ & $\emph{\textbf{i}}_{10}$  & $\emph{\textbf{i}}_{11}$ & $\emph{\textbf{i}}_{12}$  & $\emph{\textbf{i}}_{13}$ & $\emph{\textbf{i}}_{14}$ & $\emph{\textbf{i}}_{15}$
\\
$\emph{\textbf{i}}_{17}$ %
& $-\emph{\textbf{i}}_1$ & $-1$  & $-\emph{\textbf{i}}_3$ & $\emph{\textbf{i}}_2$  & $-\emph{\textbf{i}}_5$ & $\emph{\textbf{i}}_4$ & $\emph{\textbf{i}}_7$  & $-\emph{\textbf{i}}_6$
& $-\emph{\textbf{i}}_9$ & $\emph{\textbf{i}}_8$ & $\emph{\textbf{i}}_{11}$  & $-\emph{\textbf{i}}_{10}$ & $\emph{\textbf{i}}_{13}$  & $-\emph{\textbf{i}}_{12}$ & $-\emph{\textbf{i}}_{15}$ & $\emph{\textbf{i}}_{14}$
\\
$\emph{\textbf{i}}_{18}$ %
& $-\emph{\textbf{i}}_2$ & $\emph{\textbf{i}}_3$  & $-1$ & $-\emph{\textbf{i}}_1$  & $-\emph{\textbf{i}}_6$ & $-\emph{\textbf{i}}_7$ & $\emph{\textbf{i}}_4$  & $\emph{\textbf{i}}_5$
& $-\emph{\textbf{i}}_{10}$ & $-\emph{\textbf{i}}_{11}$ & $\emph{\textbf{i}}_8$  & $\emph{\textbf{i}}_9$ & $\emph{\textbf{i}}_{14}$  & $\emph{\textbf{i}}_{15}$ & $-\emph{\textbf{i}}_{12}$ & $-\emph{\textbf{i}}_{13}$
\\
$\emph{\textbf{i}}_{19}$ %
& $-\emph{\textbf{i}}_3$ & $-\emph{\textbf{i}}_2$  & $\emph{\textbf{i}}_1$ & $-1$  & $-\emph{\textbf{i}}_7$ & $\emph{\textbf{i}}_6$ & $-\emph{\textbf{i}}_5$  & $\emph{\textbf{i}}_4$
& $-\emph{\textbf{i}}_{11}$ & $\emph{\textbf{i}}_{10}$ & $-\emph{\textbf{i}}_9$  & $\emph{\textbf{i}}_8$ & $\emph{\textbf{i}}_{15}$  & $-\emph{\textbf{i}}_{14}$ & $\emph{\textbf{i}}_{13}$ & $-\emph{\textbf{i}}_{12}$
\\
\hline
$\emph{\textbf{i}}_{20}$ %
& $-\emph{\textbf{i}}_4$ & $\emph{\textbf{i}}_5$  & $\emph{\textbf{i}}_6$ & $\emph{\textbf{i}}_7$  & $-1$ & $-\emph{\textbf{i}}_1$ & $-\emph{\textbf{i}}_2$  & $-\emph{\textbf{i}}_3$
& $-\emph{\textbf{i}}_{12}$ & $-\emph{\textbf{i}}_{13}$ & $-\emph{\textbf{i}}_{14}$  & $-\emph{\textbf{i}}_{15}$ & $\emph{\textbf{i}}_{8}$  & $\emph{\textbf{i}}_{9}$ & $\emph{\textbf{i}}_{10}$ & $\emph{\textbf{i}}_{11}$
\\
$\emph{\textbf{i}}_{21}$ %
& $-\emph{\textbf{i}}_5$ & $-\emph{\textbf{i}}_4$  & $\emph{\textbf{i}}_7$ & $-\emph{\textbf{i}}_6$  & $\emph{\textbf{i}}_1$ & $-1$ & $\emph{\textbf{i}}_3$  & $-\emph{\textbf{i}}_2$
& $-\emph{\textbf{i}}_{13}$ & $\emph{\textbf{i}}_{12}$ & $-\emph{\textbf{i}}_{15}$  & $\emph{\textbf{i}}_{14}$ & $-\emph{\textbf{i}}_9$  & $\emph{\textbf{i}}_8$ & $-\emph{\textbf{i}}_{11}$ & $\emph{\textbf{i}}_{10}$
\\
$\emph{\textbf{i}}_{22}$ %
& $-\emph{\textbf{i}}_6$ & $-\emph{\textbf{i}}_7$  & $-\emph{\textbf{i}}_4$ & $\emph{\textbf{i}}_5$  & $\emph{\textbf{i}}_2$ & $-\emph{\textbf{i}}_3$ & $-1$  & $\emph{\textbf{i}}_1$
& $-\emph{\textbf{i}}_{14}$ & $\emph{\textbf{i}}_{15}$ & $\emph{\textbf{i}}_{12}$  & $-\emph{\textbf{i}}_{13}$ & $-\emph{\textbf{i}}_{10}$  & $\emph{\textbf{i}}_{11}$ & $\emph{\textbf{i}}_8$ & $-\emph{\textbf{i}}_9$
\\
$\emph{\textbf{i}}_{23}$ %
& $-\emph{\textbf{i}}_7$ & $\emph{\textbf{i}}_6$  & $-\emph{\textbf{i}}_5$ & $-\emph{\textbf{i}}_4$  & $\emph{\textbf{i}}_3$ & $\emph{\textbf{i}}_2$ & $-\emph{\textbf{i}}_1$  & $-1$
& $-\emph{\textbf{i}}_{15}$ & $-\emph{\textbf{i}}_{14}$ & $\emph{\textbf{i}}_{13}$  & $\emph{\textbf{i}}_{12}$ & $-\emph{\textbf{i}}_{11}$  & $-\emph{\textbf{i}}_{10}$ & $\emph{\textbf{i}}_9$ & $\emph{\textbf{i}}_8$
\\
\hline
$\emph{\textbf{i}}_{24}$ %
& $-\emph{\textbf{i}}_8$ & $\emph{\textbf{i}}_9$  & $\emph{\textbf{i}}_{10}$ & $\emph{\textbf{i}}_{11}$  & $\emph{\textbf{i}}_{12}$ & $\emph{\textbf{i}}_{13}$ & $\emph{\textbf{i}}_{14}$  & $\emph{\textbf{i}}_{15}$
& $-1$ & $-\emph{\textbf{i}}_1$ & $-\emph{\textbf{i}}_2$  & $-\emph{\textbf{i}}_3$ & $-\emph{\textbf{i}}_4$  & $-\emph{\textbf{i}}_5$ & $-\emph{\textbf{i}}_6$ & $-\emph{\textbf{i}}_7$
\\
$\emph{\textbf{i}}_{25}$ %
& $-\emph{\textbf{i}}_9$ & $-\emph{\textbf{i}}_8$  & $\emph{\textbf{i}}_{11}$ & $-\emph{\textbf{i}}_{10}$  & $\emph{\textbf{i}}_{13}$ & $-\emph{\textbf{i}}_{12}$ & $-\emph{\textbf{i}}_{15}$  & $\emph{\textbf{i}}_{14}$
& $\emph{\textbf{i}}_1$ & $-1$ & $\emph{\textbf{i}}_3$  & $-\emph{\textbf{i}}_2$ & $\emph{\textbf{i}}_5$  & $-\emph{\textbf{i}}_4$ & $-\emph{\textbf{i}}_7$ & $\emph{\textbf{i}}_6$
\\
$\emph{\textbf{i}}_{26}$ %
& $-\emph{\textbf{i}}_{10}$ & $-\emph{\textbf{i}}_{11}$  & $-\emph{\textbf{i}}_8$ & $\emph{\textbf{i}}_9$  & $\emph{\textbf{i}}_{14}$ & $\emph{\textbf{i}}_{15}$ & $-\emph{\textbf{i}}_{12}$  & $-\emph{\textbf{i}}_{13}$
& $\emph{\textbf{i}}_2$ & $-\emph{\textbf{i}}_3$ & $-1$  & $\emph{\textbf{i}}_1$ & $\emph{\textbf{i}}_6$  & $\emph{\textbf{i}}_7$ & $-\emph{\textbf{i}}_4$ & $-\emph{\textbf{i}}_5$
\\
$\emph{\textbf{i}}_{27}$ %
& $-\emph{\textbf{i}}_{11}$ & $\emph{\textbf{i}}_{10}$  & $-\emph{\textbf{i}}_9$ & $-\emph{\textbf{i}}_8$  & $\emph{\textbf{i}}_{15}$ & $-\emph{\textbf{i}}_{14}$ & $\emph{\textbf{i}}_{13}$  & $-\emph{\textbf{i}}_{12}$
& $\emph{\textbf{i}}_3$ & $\emph{\textbf{i}}_2$ & $-\emph{\textbf{i}}_1$  & $-1$ & $\emph{\textbf{i}}_7$  & $-\emph{\textbf{i}}_6$ & $\emph{\textbf{i}}_5$ & $-\emph{\textbf{i}}_4$
\\
\hline
$\emph{\textbf{i}}_{28}$ %
& $-\emph{\textbf{i}}_{12}$ & $-\emph{\textbf{i}}_{13}$  & $-\emph{\textbf{i}}_{14}$ & $-\emph{\textbf{i}}_{15}$  & $-\emph{\textbf{i}}_8$ & $\emph{\textbf{i}}_9$ & $\emph{\textbf{i}}_{10}$  & $\emph{\textbf{i}}_{11}$
& $\emph{\textbf{i}}_4$ & $-\emph{\textbf{i}}_5$ & $-\emph{\textbf{i}}_6$  & $-\emph{\textbf{i}}_7$ & $-1$  & $\emph{\textbf{i}}_1$ & $\emph{\textbf{i}}_2$ & $\emph{\textbf{i}}_3$
\\
$\emph{\textbf{i}}_{29}$ %
& $-\emph{\textbf{i}}_{13}$ & $\emph{\textbf{i}}_{12}$  & $-\emph{\textbf{i}}_{15}$ & $\emph{\textbf{i}}_{14}$  & $-\emph{\textbf{i}}_9$ & $-\emph{\textbf{i}}_8$ & $-\emph{\textbf{i}}_{11}$  & $\emph{\textbf{i}}_{10}$
& $\emph{\textbf{i}}_5$ & $\emph{\textbf{i}}_4$ & $-\emph{\textbf{i}}_7$  & $\emph{\textbf{i}}_6$ & $-\emph{\textbf{i}}_1$  & $-1$ & $-\emph{\textbf{i}}_3$ & $\emph{\textbf{i}}_2$
\\
$\emph{\textbf{i}}_{30}$ %
& $-\emph{\textbf{i}}_{14}$ & $\emph{\textbf{i}}_{15}$  & $\emph{\textbf{i}}_{12}$ & $-\emph{\textbf{i}}_{13}$  & $-\emph{\textbf{i}}_{10}$ & $\emph{\textbf{i}}_{11}$ & $-\emph{\textbf{i}}_8$  & $-\emph{\textbf{i}}_9$
& $\emph{\textbf{i}}_6$ & $\emph{\textbf{i}}_7$ & $\emph{\textbf{i}}_4$  & $-\emph{\textbf{i}}_5$ & $-\emph{\textbf{i}}_2$  & $\emph{\textbf{i}}_3$ & $-1$ & $-\emph{\textbf{i}}_1$
\\
$\emph{\textbf{i}}_{31}$ %
& $-\emph{\textbf{i}}_{15}$ & $-\emph{\textbf{i}}_{14}$  & $\emph{\textbf{i}}_{13}$ & $\emph{\textbf{i}}_{12}$  & $-\emph{\textbf{i}}_{11}$ & $-\emph{\textbf{i}}_{10}$ & $\emph{\textbf{i}}_9$  & $-\emph{\textbf{i}}_8$
& $\emph{\textbf{i}}_7$ & $-\emph{\textbf{i}}_6$ & $\emph{\textbf{i}}_5$  & $\emph{\textbf{i}}_4$ & $-\emph{\textbf{i}}_3$  & $-\emph{\textbf{i}}_2$ & $\emph{\textbf{i}}_1$ & $-1$
\\
\hline\hline
\end{tabular}
}
\end{table}

\section{\label{sec:level1}Trigintaduonion field equations}

In the trigintaduonion spaces, it is capable of exploring some physical quantities of four interactions, including the trigintaduonion integrating function of field potential, field potential, field strength, field source, linear momentum, angular momentum, torque, and force (Table 3). Under certain special conditions, there may be a few situations where the trigintaduonion force is equal to zero. And then it is able to deduce several continuity equations and equilibrium equations, including the fluid continuity equation, current continuity equation, and force equilibrium equation and so forth.

In the trigintaduonion spaces, the trigintaduonion integrating function of field potential can be defined as,
\begin{eqnarray}
\mathbb{X} = \mathbb{X}_g + k_{eg} \mathbb{X}_e + k_{wag} \mathbb{X}_{wa} + k_{wbg} \mathbb{X}_{wb} + k_{wcg} \mathbb{X}_{wc} + k_{sag} \mathbb{X}_{sa} + k_{sbg} \mathbb{X}_{sb} + k_{scg} \mathbb{X}_{sc} ~ ,
\end{eqnarray}
where $\mathbb{X}_g$ and $\mathbb{X}_e$ are the four-dimensional integrating functions of field potential in the subspaces, $\mathbb{H}_g$ and $\mathbb{H}_e$ , respectively, and they relate with the gravitational fields and electromagnetic fields, respectively. $\mathbb{X}_{wa}$ , $\mathbb{X}_{wb}$ , and $\mathbb{X}_{wc}$ are the four-dimensional integrating functions of field potential in the subspaces, $\mathbb{H}_{wa}$ , $\mathbb{H}_{wb}$ , and $\mathbb{H}_{wc}$ , respectively, and they are relevant to the weak interactions. $\mathbb{X}_{sa}$ , $\mathbb{X}_{sb}$ , and $\mathbb{X}_{sc}$ are the four-dimensional integrating functions of field potential in the subspaces, $\mathbb{H}_{sa}$ , $\mathbb{H}_{sb}$ , and $\mathbb{H}_{sc}$ , respectively, and they are relevant to the strong interactions. $\mathbb{X}_g = \textrm{i} \textbf{I}_{g0} X_{g0} + \Sigma \textbf{I}_{gk} X_{gk}$ . $\mathbb{X}_e = \textrm{i} \textbf{I}_{e0} X_{e0} + \Sigma \textbf{I}_{ek} X_{ek}$ . $\mathbb{X}_{wa} = \textrm{i} \textbf{I}_{wa0} X_{wa0} + \Sigma \textbf{I}_{wak} X_{wak}$ . $\mathbb{X}_{wb} = \textrm{i} \textbf{I}_{wb0} X_{wb0} + \Sigma \textbf{I}_{wbk} X_{wbk}$. $\mathbb{X}_{wc} = \textrm{i} \textbf{I}_{wc0} X_{wc0} + \Sigma \textbf{I}_{wck} X_{wck}$ . $\mathbb{X}_{sa} = \textrm{i} \textbf{I}_{sa0} X_{sa0} + \Sigma \textbf{I}_{sak} X_{sak}$. $\mathbb{X}_{sb} = \textrm{i} \textbf{I}_{sb0} X_{sb0} + \Sigma \textbf{I}_{sbk} X_{sbk}$ . $\mathbb{X}_{sc} = \textrm{i} \textbf{I}_{sc0} X_{sc0} + \Sigma \textbf{I}_{sck} X_{sck}$ . $X_{gj}$ , $X_{ej}$ , $X_{waj}$ , $X_{wbj}$ , $X_{wcj}$ , $X_{saj}$ , $X_{sbj}$ , and $X_{scj}$ are all real. $j = 0, 1, 2, 3$.

From the above, it is able to define the trigintaduonion field potential as follows,
\begin{eqnarray}
\mathbb{A} = \textrm{i} ( \lozenge + k_{xx} \mathbb{X} )^\times \circ \mathbb{X} ~ ,
\end{eqnarray}
where $k_{xx}$ is a coefficient to satisfy the needs of dimensional homogeneity. $\lozenge = \textrm{i} \textbf{I}_{g0} \partial_{g0} + \Sigma \textbf{I}_{gk} \partial_{gk}$, and $\partial_{gj} = \partial / \partial R_{gj}$. $\times$ denotes the complex conjugate. The symbol $\circ$ indicates the trigintaduonion multiplication. $j = 0, 1, 2, 3$.

In the above, the trigintaduonion field potential can be further rewritten as,
\begin{eqnarray}
\mathbb{A} = \mathbb{A}_g + k_{eg} \mathbb{A}_e + k_{wag} \mathbb{A}_{wa} + k_{wbg} \mathbb{A}_{wb} + k_{wcg} \mathbb{A}_{wc} + k_{sag} \mathbb{A}_{sa} + k_{sbg} \mathbb{A}_{sb} + k_{scg} \mathbb{A}_{sc} ~ ,
\end{eqnarray}
where $\mathbb{A}_g$ and $\mathbb{A}_e$ are the four-dimensional field potentials in the subspaces, $\mathbb{H}_g$ and $\mathbb{H}_e$ , respectively, and they relate with the gravitational fields and electromagnetic fields, respectively. $\mathbb{A}_{wa}$ , $\mathbb{A}_{wb}$ , and $\mathbb{A}_{wc}$ are the four-dimensional field potentials in the subspaces, $\mathbb{H}_{wa}$ , $\mathbb{H}_{wb}$ , and $\mathbb{H}_{wc}$ , respectively, and they are relevant to the weak interactions. $\mathbb{A}_{sa}$, $\mathbb{A}_{sb}$ , and $\mathbb{A}_{sc}$ are the four-dimensional field potentials in the subspaces, $\mathbb{H}_{sa}$ , $\mathbb{H}_{sb}$ , and $\mathbb{H}_{sc}$ , respectively, and they are relevant to the strong interactions. Under some gauge conditions, the field potential can be written in a simple form, such as, $\mathbb{A}_g = \textrm{i} \textbf{I}_{g0} A_{g0} + \Sigma \textbf{I}_{gk} A_{gk}$. $\mathbb{A}_e = \textrm{i} \textbf{I}_{e0} A_{e0} + \Sigma \textbf{I}_{ek} A_{ek}$ . $\mathbb{A}_{wa} = \textrm{i} \textbf{I}_{wa0} A_{wa0} + \Sigma \textbf{I}_{wak} A_{wak}$. $\mathbb{A}_{wb} = \textrm{i} \textbf{I}_{wb0} A_{wb0} + \Sigma \textbf{I}_{wbk} A_{wbk}$. $\mathbb{A}_{wc} = \textrm{i} \textbf{I}_{wc0} A_{wc0} + \Sigma \textbf{I}_{wck} A_{wck}$ . $\mathbb{A}_{sa} = \textrm{i} \textbf{I}_{sa0} A_{sa0} + \Sigma \textbf{I}_{sak} A_{sak}$. $\mathbb{A}_{sb} = \textrm{i} \textbf{I}_{sb0} A_{sb0} + \Sigma \textbf{I}_{sbk} A_{sbk}$. $\mathbb{A}_{sc} = \textrm{i} \textbf{I}_{sc0} A_{sc0} + \Sigma \textbf{I}_{sck} A_{sck}$ . $A_{gj}$ , $A_{ej}$ , $A_{waj}$ , $A_{wbj}$ , $A_{wcj}$ , $A_{saj}$ , $A_{sbj}$ , and $A_{scj}$ are all real. $j = 0, 1, 2, 3$.

It is able to define the trigintaduonion field strength from two equations in the above,
\begin{eqnarray}
\mathbb{F} = ( \lozenge + k_{ax} \mathbb{X} + k_{aa} \mathbb{A} ) \circ \mathbb{A} ~ ,
\end{eqnarray}
where $k_{ax}$ and $k_{aa}$ are coefficients, to meet the requirement of dimensional homogeneity.

In the above, the trigintaduonion field strength can be further rewritten as,
\begin{eqnarray}
\mathbb{F} = \mathbb{F}_g + k_{eg} \mathbb{F}_e + k_{wag} \mathbb{F}_{wa} + k_{wbg} \mathbb{F}_{wb} + k_{wcg} \mathbb{F}_{wc} + k_{sag} \mathbb{F}_{sa} + k_{sbg} \mathbb{F}_{sb} + k_{scg} \mathbb{F}_{sc} ~ ,
\end{eqnarray}
where $\mathbb{F}_g$ and $\mathbb{F}_e$ are the four-dimensional field strengths in the subspaces, $\mathbb{H}_g$ and $\mathbb{H}_e$ , respectively, and they relate with the gravitational fields and electromagnetic fields, respectively. $\mathbb{F}_{wa}$ , $\mathbb{F}_{wb}$ , and $\mathbb{F}_{wc}$ are the four-dimensional field strengths in the subspaces, $\mathbb{H}_{wa}$ , $\mathbb{H}_{wb}$ , and $\mathbb{H}_{wc}$ , respectively, and they are relevant to the weak interactions. $\mathbb{F}_{sa}$ , $\mathbb{F}_{sb}$ , and $\mathbb{F}_{sc}$ are the four-dimensional field strengths in the subspaces, $\mathbb{H}_{sa}$ , $\mathbb{H}_{sb}$ , and $\mathbb{H}_{sc}$ , respectively, and they are relevant to the strong interactions. $\mathbb{F}_g = \textbf{I}_{g0} F_{g0} + \Sigma \textbf{I}_{gk} F_{gk}$. $\mathbb{F}_e = \textbf{I}_{e0} F_{e0} + \Sigma \textbf{I}_{ek} F_{ek}$ . $\mathbb{F}_{wa} = \textbf{I}_{wa0} F_{wa0} + \Sigma \textbf{I}_{wak} F_{wak}$. $\mathbb{F}_{wb} = \textbf{I}_{wb0} F_{wb0} + \Sigma \textbf{I}_{wbk} F_{wbk}$ . $\mathbb{F}_{wc} = \textbf{I}_{wc0} F_{wc0} + \Sigma \textbf{I}_{wck} F_{wck}$. $\mathbb{F}_{sa} = \textbf{I}_{sa0} F_{sa0} + \Sigma \textbf{I}_{sak} F_{sak}$ . $\mathbb{F}_{sb} = \textbf{I}_{sb0} F_{sb0} + \Sigma \textbf{I}_{sbk} F_{sbk}$. $\mathbb{F}_{sc} = \textbf{I}_{sc0} F_{sc0} + \Sigma \textbf{I}_{sck} F_{sck}$ . $F_{g0}$ , $F_{e0}$ , $F_{wa0}$ , $F_{wb0}$ , $F_{wc0}$ , $F_{sa0}$ , $F_{sb0}$, and $F_{sc0}$ are all real. $F_{gk}$ , $F_{ek}$ , $F_{wak}$ , $F_{wbk}$, $F_{wck}$ , $F_{sak}$ , $F_{sbk}$ , and $F_{sck}$ are the complex numbers. In particular, when $F_{g0}$ , $F_{e0}$, $F_{wa0}$ , $F_{wb0}$ , $F_{wc0}$ , $F_{sa0}$ , $F_{sb0}$ , and $F_{sc0}$ are all equal to zero, it is capable of achieving a few gauge conditions, including the Lorentz gauge condition in the electromagnetic fields and others. $j = 0, 1, 2, 3$.

Similarly, it is capable of defining the trigintaduonion field source,
\begin{eqnarray}
\mu \mathbb{S} = - ( \lozenge + k_{fx} \mathbb{X} + k_{fa} \mathbb{A} + k_{ff} \mathbb{F})^\ast \circ \mathbb{F} ~ ,
\end{eqnarray}
where $\mu$ , $k_{fx}$ , $k_{fa}$ , and $k_{ff}$ are the coefficients, to satisfy the needs of dimensional homogeneity. $\ast$ denotes the trigintaduonion conjugate.

In the above, the trigintaduonion field source can be further rewritten as,
\begin{eqnarray}
\mu \mathbb{S} = \mu_g \mathbb{S}_g + k_{eg} \mu_e \mathbb{S}_e + k_{wag} \mu_{wa} \mathbb{S}_{wa} + k_{wbg} \mu_{wb} \mathbb{S}_{wb} + k_{wcg} \mu_{wc} \mathbb{S}_{wc} + k_{sag} \mu_{sa} \mathbb{S}_{sa} + k_{sbg} \mu_{sb} \mathbb{S}_{sb} + k_{scg} \mu_{sc} \mathbb{S}_{sc} ~ ,
\end{eqnarray}
where $\mathbb{S}_g$ and $\mathbb{S}_e$ are the four-dimensional field sources in the subspaces, $\mathbb{H}_g$ and $\mathbb{H}_e$ , respectively, and they relate with the gravitational fields and electromagnetic fields, respectively. $\mathbb{S}_{wa}$ , $\mathbb{S}_{wb}$ , and $\mathbb{S}_{wc}$ are the four-dimensional field sources in the subspaces, $\mathbb{H}_{wa}$ , $\mathbb{H}_{wb}$ , and $\mathbb{H}_{wc}$ , respectively, and they are relevant to the weak interactions. $\mathbb{S}_{sa}$ , $\mathbb{S}_{sb}$ , and $\mathbb{S}_{sc}$ are the four-dimensional field sources in the subspaces, $\mathbb{H}_{sa}$ , $\mathbb{H}_{sb}$ , and $\mathbb{H}_{sc}$ , respectively, and they are relevant to the strong interactions. $\mu_g$, $\mu_e$, $\mu_{wa}$, $\mu_{wb}$ , $\mu_{wc}$ , $\mu_{sa}$ , $\mu_{sb}$ , and $\mu_{sc}$ are coefficients, to meet the requirement of dimensional homogeneity. $\mathbb{S}_g = \textrm{i} \textbf{I}_{g0} S_{g0} + \Sigma \textbf{I}_{gk} S_{gk}$. $\mathbb{S}_e = \textrm{i} \textbf{I}_{e0} S_{e0} + \Sigma \textbf{I}_{ek} S_{ek}$ . $\mathbb{S}_{wa} = \textrm{i} \textbf{I}_{wa0} S_{wa0} + \Sigma \textbf{I}_{wak} S_{wak}$. $\mathbb{S}_{wb} = \textrm{i} \textbf{I}_{wb0} S_{wb0} + \Sigma \textbf{I}_{wbk} S_{wbk}$ . $\mathbb{S}_{wc} = \textrm{i} \textbf{I}_{wc0} S_{wc0} + \Sigma \textbf{I}_{wck} S_{wck}$. $\mathbb{S}_{sa} = \textrm{i} \textbf{I}_{sa0} S_{sa0} + \Sigma \textbf{I}_{sak} S_{sak}$. $\mathbb{S}_{sb} = \textrm{i} \textbf{I}_{sb0} S_{sb0} + \Sigma \textbf{I}_{sbk} S_{sbk}$. $\mathbb{S}_{sc} = \textrm{i} \textbf{I}_{sc0} S_{sc0} + \Sigma \textbf{I}_{sck} S_{sck}$ . $(\textbf{I}_{g0} S_{g0} / v_0)$ is the inertial mass. $\Sigma \textbf{I}_{gk} S_{gk}$ is the linear momentum. $(\textbf{I}_{e0} S_{e0} / v_0)$ is the electric charge. $\Sigma \textbf{I}_{ek} S_{ek}$ is the electric current. $S_{gj}$ , $S_{ej}$ , $S_{waj}$ , $S_{wbj}$ , $S_{wcj}$ , $S_{saj}$ , $S_{sbj}$ , and $S_{scj}$ are all real. Specifically, by means of the variable separation methods, some field equations can be obtained from the above two equations, including the electromagnetic equations. $j = 0, 1, 2, 3$.

\begin{table}[h]
\caption{The quaternion operator and various trigintaduonion physical quantities can be combined together to constitute a few multiple composite operators, achieving some trigintaduonion physical quantities that are related to the contributions of different physical quantities.}
\center
\begin{tabular}{@{}ll@{}}
\hline\hline
physical quantity                      &   definition                                                                                               \\
\hline
trigintaduonion field potential        &   $\mathbb{A} = \textrm{i} ( \lozenge + k_{xx} \mathbb{X} )^\times \circ \mathbb{X}$                                \\
trigintaduonion field strength         &   $\mathbb{F} = ( \lozenge + k_{ax} \mathbb{X} + k_{aa} \mathbb{A} ) \circ \mathbb{A}$                     \\
trigintaduonion field source           &   $\mu \mathbb{S} = - ( \lozenge + k_{fx} \mathbb{X} + k_{fa} \mathbb{A} + k_{ff} \mathbb{F})^\ast \circ \mathbb{F}$   \\
trigintaduonion linear momentum        &   $\mathbb{P} = \mu \mathbb{S} / \mu_g$                                                                    \\
trigintaduonion angular momentum       &   $\mathbb{L} = ( \mathbb{R} + k_{rx} \mathbb{X} )^\times \circ \mathbb{P}$                                \\
trigintaduonion torque                 &   $\mathbb{W} = - v_0 ( \lozenge + k_{lx} \mathbb{X} + k_{la} \mathbb{A} + k_{lf} \mathbb{F} + k_{ll} \mathbb{L} ) \circ \{ ( i \mathbb{V}^\times / v_0 ) \circ \mathbb{L} \} $
\\
trigintaduonion force                  &   $\mathbb{N} = - ( \lozenge + k_{wx} \mathbb{X} + k_{wa} \mathbb{A} + k_{wf} \mathbb{F} + k_{wl} \mathbb{L} + k_{ww} \mathbb{W} ) \circ \{ ( i \mathbb{V}^\times / v_0 ) \circ \mathbb{W} \}  $
\\
\hline\hline
\end{tabular}
\end{table}

In the trigintaduonion spaces, the trigintaduonion linear momentum can be written as,
\begin{eqnarray}
\mathbb{P} = \mu \mathbb{S} / \mu_g ~ ,
\end{eqnarray}
or
\begin{eqnarray}
\mathbb{P} = \mathbb{P}_g + k_{eg} \mathbb{P}_e + k_{wag} \mathbb{P}_{wa} + k_{wbg} \mathbb{P}_{wb} + k_{wcg} \mathbb{P}_{wc} + k_{sag} \mathbb{P}_{sa} + k_{sbg} \mathbb{P}_{sb} + k_{scg} \mathbb{P}_{sc} ~ ,
\end{eqnarray}
where $\mathbb{P}_g$ and $\mathbb{P}_e$ are the four-dimensional linear momenta in the subspaces, $\mathbb{H}_g$ and $\mathbb{H}_e$ , respectively, and they relate with the gravitational fields and electromagnetic fields, respectively. $\mathbb{P}_{wa}$ , $\mathbb{P}_{wb}$ , and $\mathbb{P}_{wc}$ are the four-dimensional linear momenta in the subspaces, $\mathbb{H}_{wa}$ , $\mathbb{H}_{wb}$ , and $\mathbb{H}_{wc}$ , respectively, and they are relevant to the weak interactions. $\mathbb{P}_{sa}$ , $\mathbb{P}_{sb}$ , and $\mathbb{P}_{sc}$ are the four-dimensional linear momenta in the subspaces, $\mathbb{H}_{sa}$ , $\mathbb{H}_{sb}$ , and $\mathbb{H}_{sc}$ , respectively, and they are relevant to the strong interactions. $\mathbb{P}_g = \textrm{i} \textbf{I}_{g0} P_{g0} + \Sigma \textbf{I}_{gk} P_{gk}$ . $\mathbb{P}_e = \textrm{i} \textbf{I}_{e0} P_{e0} + \Sigma \textbf{I}_{ek} P_{ek}$ . $\mathbb{P}_{wa} = \textrm{i} \textbf{I}_{wa0} P_{wa0} + \Sigma \textbf{I}_{wak} P_{wak}$. $\mathbb{P}_{wb} = \textrm{i} \textbf{I}_{wb0} P_{wb0} + \Sigma \textbf{I}_{wbk} P_{wbk}$ . $\mathbb{P}_{wc} = \textrm{i} \textbf{I}_{wc0} P_{wc0} + \Sigma \textbf{I}_{wck} P_{wck}$ . $\mathbb{P}_{sa} = \textrm{i} \textbf{I}_{sa0} P_{sa0} + \Sigma \textbf{I}_{sak} P_{sak}$. $\mathbb{P}_{sb} = \textrm{i} \textbf{I}_{sb0} P_{sb0} + \Sigma \textbf{I}_{sbk} P_{sbk}$. $\mathbb{P}_{sc} = \textrm{i} \textbf{I}_{sc0} P_{sc0} + \Sigma \textbf{I}_{sck} P_{sck}$ . $P_{gj}$ , $P_{ej}$ , $P_{waj}$ , $P_{wbj}$ , $P_{wcj}$ , $P_{saj}$ , $P_{sbj}$ , and $P_{scj}$ are all real. $j = 0, 1, 2, 3$.

From the above, the trigintaduonion angular momentum can be written as,
\begin{eqnarray}
\mathbb{L} = ( \mathbb{R} + k_{rx} \mathbb{X} )^\times \circ \mathbb{P} ~ ,
\end{eqnarray}
or
\begin{eqnarray}
\mathbb{L} = \mathbb{L}_g + k_{eg} \mathbb{L}_e + k_{wag} \mathbb{L}_{wa} + k_{wbg} \mathbb{L}_{wb} + k_{wcg} \mathbb{L}_{wc} + k_{sag} \mathbb{L}_{sa} + k_{sbg} \mathbb{L}_{sb} + k_{scg} \mathbb{L}_{sc} ~ ,
\end{eqnarray}
where $\mathbb{L}_g$ and $\mathbb{L}_e$ are the four-dimensional angular momenta in the subspaces, $\mathbb{H}_g$ and $\mathbb{H}_e$ , respectively, and they relate with the gravitational fields and electromagnetic fields, respectively. $\mathbb{L}_{wa}$ , $\mathbb{L}_{wb}$ , and $\mathbb{L}_{wc}$ are the four-dimensional angular momenta in the subspaces, $\mathbb{H}_{wa}$ , $\mathbb{H}_{wb}$ , and $\mathbb{H}_{wc}$ , respectively, and they are relevant to the weak interactions. $\mathbb{L}_{sa}$ , $\mathbb{L}_{sb}$ , and $\mathbb{L}_{sc}$ are the four-dimensional angular momenta in the subspaces, $\mathbb{H}_{sa}$ , $\mathbb{H}_{sb}$ , and $\mathbb{H}_{sc}$ , respectively, and they are relevant to the strong interactions. $k_{rx}$ is one coefficient, to satisfy the needs of dimensional homogeneity. $\mathbb{L}_g = \textbf{I}_{g0} L_{g0} + \textrm{i} \Sigma \textbf{I}_{gk} L_{gk}^i + \Sigma \textbf{I}_{gk} L_{gk}$ . $\mathbb{L}_e = \textbf{I}_{e0} L_{e0} + \textrm{i} \Sigma \textbf{I}_{ek} L_{ek}^i + \Sigma \textbf{I}_{ek} L_{ek}$ . $\mathbb{L}_{wa} = \textbf{I}_{wa0} L_{wa0} + \textrm{i} \Sigma \textbf{I}_{wak} L_{wak}^i + \Sigma \textbf{I}_{wak} L_{wak}$. $\mathbb{L}_{wb} = \textbf{I}_{wb0} L_{wb0} + \textrm{i} \Sigma \textbf{I}_{wbk} L_{wbk}^i + \Sigma \textbf{I}_{wbk} L_{wbk}$ . $\mathbb{L}_{wc} = \textbf{I}_{wc0} L_{wc0} + \textrm{i} \Sigma \textbf{I}_{wck} L_{wck}^i + \Sigma \textbf{I}_{wck} L_{wck}$ . $\mathbb{L}_{sa} = \textbf{I}_{sa0} L_{sa0} + \textrm{i} \Sigma \textbf{I}_{sak} L_{sak}^i + \Sigma \textbf{I}_{sak} L_{sak}$. $\mathbb{L}_{sb} = \textbf{I}_{sb0} L_{sb0} + \textrm{i} \Sigma \textbf{I}_{sbk} L_{sbk}^i + \Sigma \textbf{I}_{sbk} L_{sbk}$. $\mathbb{L}_{sc} = \textbf{I}_{sc0} L_{sc0} + \textrm{i} \Sigma \textbf{I}_{sck} L_{sck}^i + \Sigma \textbf{I}_{sck} L_{sck}$ . $\Sigma \textbf{I}_{gk} L_{gk}$ is the angular momentum. $\Sigma \textbf{I}_{ek} L_{ek}^i$ is the electric moment. $\Sigma \textbf{I}_{ek} L_{ek}$ is the magnetic moment. $L_{gj}$ , $L_{ej}$ , $L_{waj}$ , $L_{wbj}$, $L_{wcj}$ , $L_{saj}$ , $L_{sbj}$ , $L_{scj}$ , $L_{gk}^i$ , $L_{ek}^i$ , $L_{wak}^i$ , $L_{wbk}^i$ , $L_{wck}^i$ , $L_{sak}^i$ , $L_{sbk}^i$ , and $L_{sck}^i$ are all real. $j = 0, 1, 2, 3$.

The trigintaduonion torque can be written as,
\begin{eqnarray}
\mathbb{W} = - v_0 ( \lozenge + k_{lx} \mathbb{X} + k_{la} \mathbb{A} + k_{lf} \mathbb{F} + k_{ll} \mathbb{L} ) \circ \{ ( i \mathbb{V}^\times / v_0 ) \circ \mathbb{L} \}  ~ ,
\end{eqnarray}
or
\begin{eqnarray}
\mathbb{W} = \mathbb{W}_g + k_{eg} \mathbb{W}_e + k_{wag} \mathbb{W}_{wa} + k_{wbg} \mathbb{W}_{wb} + k_{wcg} \mathbb{W}_{wc} + k_{sag} \mathbb{W}_{sa} + k_{sbg} \mathbb{W}_{sb} + k_{scg} \mathbb{W}_{sc} ~ ,
\end{eqnarray}
where $\mathbb{V} = \partial \mathbb{R}  / \partial t $ is the trigintaduonion velocity. $\mathbb{W}_g$ and $\mathbb{W}_e$ are the four-dimensional torques in the subspaces, $\mathbb{H}_g$ and $\mathbb{H}_e$ , respectively, and they relate with the gravitational fields and electromagnetic fields, respectively. $\mathbb{W}_{wa}$ , $\mathbb{W}_{wb}$ , and $\mathbb{W}_{wc}$ are the four-dimensional torques in the subspaces, $\mathbb{H}_{wa}$ , $\mathbb{H}_{wb}$, and $\mathbb{H}_{wc}$ , respectively, and they are relevant to the weak interactions. $\mathbb{W}_{sa}$, $\mathbb{W}_{sb}$ , and $\mathbb{W}_{sc}$ are the four-dimensional torques in the subspaces, $\mathbb{H}_{sa}$ , $\mathbb{H}_{sb}$ , and $\mathbb{H}_{sc}$ , respectively, and they are relevant to the strong interactions. $k_{lx}$ , $k_{la}$ , $k_{lf}$ , and $k_{ll}$ are coefficients, to meet the requirement of dimensional homogeneity. $\mathbb{W}_g = \textrm{i} \textbf{I}_{g0} W_{g0}^i + \textbf{I}_{g0} W_{g0} + \textrm{i} \Sigma \textbf{I}_{gk} W_{gk}^i + \Sigma \textbf{I}_{gk} W_{gk}$ . $\mathbb{W}_e = \textrm{i} \textbf{I}_{e0} W_{e0}^i + \textbf{I}_{e0} W_{e0} + \textrm{i} \Sigma \textbf{I}_{ek} W_{ek}^i + \Sigma \textbf{I}_{ek} W_{ek}$ . $\mathbb{W}_{wa} = \textrm{i} \textbf{I}_{wa0} W_{wa0}^i + \textbf{I}_{wa0} W_{wa0} + \textrm{i} \Sigma \textbf{I}_{wak} W_{wak}^i + \Sigma \textbf{I}_{wak} W_{wak}$ . $\mathbb{W}_{wb} = \textrm{i} \textbf{I}_{wb0} W_{wb0}^i + \textbf{I}_{wb0} W_{wb0} + \textrm{i} \Sigma \textbf{I}_{wbk} W_{wbk}^i + \Sigma \textbf{I}_{wbk} W_{wbk}$. $\mathbb{W}_{wc} = \textrm{i} \textbf{I}_{wc0} W_{wc0}^i + \textbf{I}_{wc0} W_{wc0} + \textrm{i} \Sigma \textbf{I}_{wck} W_{wck}^i + \Sigma \textbf{I}_{wck} W_{wck}$. $\mathbb{W}_{sa} = \textrm{i} \textbf{I}_{sa0} W_{sa0}^i + \textbf{I}_{sa0} W_{sa0} + \textrm{i} \Sigma \textbf{I}_{sak} W_{sak}^i + \Sigma \textbf{I}_{sak} W_{sak}$. $\mathbb{W}_{sb} = \textrm{i} \textbf{I}_{sb0} W_{sb0}^i + \textbf{I}_{sb0} W_{sb0} + \textrm{i} \Sigma \textbf{I}_{sbk} W_{sbk}^i + \Sigma \textbf{I}_{sbk} W_{sbk}$ . $\mathbb{W}_{sc} = \textrm{i} \textbf{I}_{sc0} W_{sc0}^i + \textbf{I}_{sc0} W_{sc0} + \textrm{i} \Sigma \textbf{I}_{sck} W_{sck}^i + \Sigma \textbf{I}_{sck} W_{sck}$ . $\textbf{I}_{g0} W_{g0}^i$ is the energy. $\Sigma \textbf{I}_{gk} W_{gk}^i$ is the torque, including the gyroscopic torque. $W_{gj}$ , $W_{ej}$ , $W_{waj}$ , $W_{wbj}$ , $W_{wcj}$ , $W_{saj}$ , $W_{sbj}$ , $W_{scj}$ , $W_{gj}^i$ , $W_{ej}^i$ , $W_{waj}^i$ , $W_{wbj}^i$ , $W_{wcj}^i$ , $W_{saj}^i$ , $W_{sbj}^i$ , and $W_{scj}^i$ are all real. $j = 0, 1, 2, 3$.

The trigintaduonion force can be written as,
\begin{eqnarray}
\mathbb{N} = - ( \lozenge + k_{wx} \mathbb{X} + k_{wa} \mathbb{A} + k_{wf} \mathbb{F} + k_{wl} \mathbb{L} + k_{ww} \mathbb{W} ) \circ \{ ( i \mathbb{V}^\times / v_0 ) \circ \mathbb{W} \}  ~ ,
\end{eqnarray}
or
\begin{eqnarray}
\mathbb{N} = \mathbb{N}_g + k_{eg} \mathbb{N}_e + k_{wag} \mathbb{N}_{wa} + k_{wbg} \mathbb{N}_{wb} + k_{wcg} \mathbb{N}_{wc} + k_{sag} \mathbb{N}_{sa} + k_{sbg} \mathbb{N}_{sb} + k_{scg} \mathbb{N}_{sc} ~ ,
\end{eqnarray}
where $\mathbb{N}_g$ and $\mathbb{N}_e$ are the four-dimensional forces in the subspaces, $\mathbb{H}_g$ and $\mathbb{H}_e$ , respectively, and they relate with the gravitational fields and electromagnetic fields, respectively. $\mathbb{N}_{wa}$ , $\mathbb{N}_{wb}$ , and $\mathbb{N}_{wc}$ are the four-dimensional forces in the subspaces, $\mathbb{H}_{wa}$ , $\mathbb{H}_{wb}$ , and $\mathbb{H}_{wc}$, respectively, and they are relevant to the weak interactions. $\mathbb{N}_{sa}$ , $\mathbb{N}_{sb}$ , and $\mathbb{N}_{sc}$ are the four-dimensional forces in the subspaces, $\mathbb{H}_{sa}$ , $\mathbb{H}_{sb}$ , and $\mathbb{H}_{sc}$ , respectively, and they are relevant to the strong interactions. $k_{wx}$ , $k_{wa}$ , $k_{wf}$ , $k_{wl}$ , and $k_{ww}$ are coefficients, to satisfy the needs of dimensional homogeneity. $\mathbb{N}_g = \textrm{i} \textbf{I}_{g0} N_{g0}^i + \textbf{I}_{g0} N_{g0} + \textrm{i} \Sigma \textbf{I}_{gk} N_{gk}^i + \Sigma \textbf{I}_{gk} N_{gk}$ . $\mathbb{N}_e = \textrm{i} \textbf{I}_{e0} N_{e0}^i + \textbf{I}_{e0} N_{e0} + \textrm{i} \Sigma \textbf{I}_{ek} N_{ek}^i + \Sigma \textbf{I}_{ek} N_{ek}$ . $\mathbb{N}_{wa} = \textrm{i} \textbf{I}_{wa0} N_{wa0}^i + \textbf{I}_{wa0} N_{wa0} + \textrm{i} \Sigma \textbf{I}_{wak} N_{wak}^i + \Sigma \textbf{I}_{wak} N_{wak}$ . $\mathbb{N}_{wb} = \textrm{i} \textbf{I}_{wb0} N_{wb0}^i + \textbf{I}_{wb0} N_{wb0} + \textrm{i} \Sigma \textbf{I}_{wbk} N_{wbk}^i + \Sigma \textbf{I}_{wbk} N_{wbk}$. $\mathbb{N}_{wc} = \textrm{i} \textbf{I}_{wc0} N_{wc0}^i + \textbf{I}_{wc0} N_{wc0} + \textrm{i} \Sigma \textbf{I}_{wck} N_{wck}^i + \Sigma \textbf{I}_{wck} N_{wck}$ . $\mathbb{N}_{sa} = \textrm{i} \textbf{I}_{sa0} N_{sa0}^i + \textbf{I}_{sa0} N_{sa0} + \textrm{i} \Sigma \textbf{I}_{sak} N_{sak}^i + \Sigma \textbf{I}_{sak} N_{sak}$. $\mathbb{N}_{sb} = \textrm{i} \textbf{I}_{sb0} N_{sb0}^i + \textbf{I}_{sb0} N_{sb0} + \textrm{i} \Sigma \textbf{I}_{sbk} N_{sbk}^i + \Sigma \textbf{I}_{sbk} N_{sbk}$ . $\mathbb{N}_{sc} = \textrm{i} \textbf{I}_{sc0} N_{sc0}^i + \textbf{I}_{sc0} N_{sc0} + \textrm{i} \Sigma \textbf{I}_{sck} N_{sck}^i + \Sigma \textbf{I}_{sck} N_{sck}$. $\textbf{I}_{g0} N_{g0}$ is the power. $\Sigma \textbf{I}_{gk} N_{gk}^i$ is the force, including the Magnus force. $N_{gj}$ , $N_{ej}$ , $N_{waj}$ , $N_{wbj}$ , $N_{wcj}$ , $N_{saj}$ , $N_{sbj}$ , $N_{scj}$ , $N_{gj}^i$, $N_{ej}^i$ , $N_{waj}^i$ , $N_{wbj}^i$ , $N_{wcj}^i$ , $N_{saj}^i$ , $N_{sbj}^i$ , and $N_{scj}^i$ are all real. $j = 0, 1, 2, 3$.

In case $\mathbb{N} = 0$ under certain conditions, it is able to achieve thirty-two continuity equations and equilibrium equations independent of each other, including the fluid continuity equation, current continuity equation, force equilibrium equation, torque continuity equation (see Ref.[29]), second-torque continuity equation (see Ref.[30]), second-force equilibrium equation (see Ref.[31]), precession equilibrium equation \cite{weng4}, and second-precession equilibrium equation \cite{weng5}.

\section{\label{sec:level1}Gauge fields}

From the perspective of multiple quaternion spaces, the electroweak theory reveals that the weak nuclear fields are composed of three fundamental fields. These three fundamental fields are different from the electromagnetic fields, gravitational fields, or strong nuclear fields. The weak nuclear field involves three quaternion spaces independent of each other. This means that the spaces involved in weak nuclear fields have been expanded from four dimensions to twelve dimensions.

The electroweak theory involves four fundamental fields. In a larger unified theory, if we expect to include the gravitational fields, the number of fundamental fields needs to be expanded from four to five. And these five fundamental fields require five quaternion spaces independent of each other. However, these five independent quaternion spaces are incapable of satisfying the needs of multiplication table, according to the multiplication table of multidimensional spaces related to the quaternion spaces. Consequently, the minimum number of independent quaternion spaces is eight for this multidimensional space, that is, the trigintaduonion space. Eight independent quaternion spaces are able to satisfy the needs of multiplication table of this multidimensional space. This means that the strong nuclear field is also composed of three fundamental fields.

(1) In the trigintaduonion spaces, when there are $\mathbb{X} = 0$, $\mathbb{A}_e = 0$, $\mathbb{A}_g = 0$, $\mathbb{A}_{sa} = 0$, $\mathbb{A}_{sb} = 0$, and $\mathbb{A}_{sc} = 0$ simultaneously, the trigintaduonion field strength,
\begin{eqnarray}
\mathbb{F} = && \{ \lozenge + k_{aa} ( k_{wag} \mathbb{A}_{wa} + k_{wbg} \mathbb{A}_{wb} + k_{wcg} \mathbb{A}_{wc} ) \}
\nonumber
\\
&&
~~~
\circ ( k_{wag} \mathbb{A}_{wa} + k_{wbg} \mathbb{A}_{wb} + k_{wcg} \mathbb{A}_{wc} ) ~ ,
\end{eqnarray}
can be written as,
\begin{eqnarray}
\mathbb{F} =
&& \lozenge \circ ( k_{wag} \mathbb{A}_{wa} + k_{wbg} \mathbb{A}_{wb} + k_{wcg} \mathbb{A}_{wc} )
\nonumber
\\
&&
~~~
+ k_{aa} ( k_{wag}^2 \mathbb{A}_{wa} \circ \mathbb{A}_{wa} + k_{wbg}^2 \mathbb{A}_{wb} \circ \mathbb{A}_{wb} + k_{wcg}^2 \mathbb{A}_{wc} \circ \mathbb{A}_{wc} )
\nonumber
\\
&&
~~~
+ k_{aa} k_{wag} k_{wbg} ( \mathbb{A}_{wa} \circ \mathbb{A}_{wb} + \mathbb{A}_{wb} \circ \mathbb{A}_{wa} )
\nonumber
\\
&&
~~~
+ k_{aa} k_{wag} k_{wcg} ( \mathbb{A}_{wa} \circ \mathbb{A}_{wc} + \mathbb{A}_{wc} \circ \mathbb{A}_{wa} )
\nonumber
\\
&&
~~~
+ k_{aa} k_{wbg} k_{wcg} ( \mathbb{A}_{wb} \circ \mathbb{A}_{wc} + \mathbb{A}_{wc} \circ \mathbb{A}_{wb} ) ~ .
\end{eqnarray}

Due to the fact that these three field potentials, $\mathbb{A}_{wa}$ , $\mathbb{A}_{wb}$ , $\mathbb{A}_{wc}$ , of weak interactions are independent of each other, the above equation is consistent with the definition of field strength in the Yang-Mills gauge field.

In the trigintaduonion spaces, when $\mathbb{X} = 0$ and only considering the contributions of $\mathbb{A}_{wa}$ , $\mathbb{A}_{wb}$ , $\mathbb{A}_{wc}$ , $\mathbb{F}_{wa}$ , $\mathbb{F}_{wb}$ , and $\mathbb{F}_{wc}$ , the trigintaduonion field source of weak interactions can be written as,
\begin{eqnarray}
\mu \mathbb{S} = && - \{ \lozenge + k_{fa} ( k_{wag} \mathbb{A}_{wa} + k_{wbg} \mathbb{A}_{wb} + k_{wcg} \mathbb{A}_{wc} )
\nonumber
\\
&&
~~~
~~~
+ k_{ff} ( k_{wag} \mathbb{F}_{wa} + k_{wbg} \mathbb{F}_{wb} + k_{wcg} \mathbb{F}_{wc} ) \}^\ast
\nonumber
\\
&&
~~~
~~~
~~~
\circ ( k_{wag} \mathbb{F}_{wa} + k_{wbg} \mathbb{F}_{wb} + k_{wcg} \mathbb{F}_{wc} ) ~ .
\end{eqnarray}

Due to these three field strengths, $\mathbb{F}_{wa}$ , $\mathbb{F}_{wb}$ , $\mathbb{F}_{wc}$ , of weak interactions being independent of each other, the above equation is consistent with the definition of field source in the Yang-Mills gauge field \cite{izergin}.

(2) In the trigintaduonion spaces, there are four similar relational expressions, those are, $\textbf{I}_{ej} = \textbf{I}_{gj} \circ \textbf{I}_{e0}$ , $\textbf{I}_{waj} = \textbf{I}_{gj} \circ \textbf{I}_{wa0}$ , $\textbf{I}_{wbj} = \textbf{I}_{gj} \circ \textbf{I}_{wb0}$ , and $\textbf{I}_{wcj} = \textbf{I}_{gj} \circ \textbf{I}_{wc0}$ . Therefore, it is able to combine the electromagnetic potential and weak nuclear potential into one simpler form. When there are $\mathbb{A}_g = 0$, $\mathbb{A}_{sa} = 0$, $\mathbb{A}_{sb} = 0$, and $\mathbb{A}_{sc} = 0$ simultaneously, the trigintaduonion field potential can be simplified into,
\begin{eqnarray}
\mathbb{A} = k_{eg} \mathbb{A}_e + k_{wag} \mathbb{A}_{wa} + k_{wbg} \mathbb{A}_{wb} + k_{wcg} \mathbb{A}_{wc} ~ .
\end{eqnarray}

The above can be rewritten as,
\begin{eqnarray}
\mathbb{A} = k_{eg} \mathbb{A}_e + k_{wag} k_{sa}^\prime \textbf{I}_{sa}^\prime \circ \mathbb{A}_{wa}^\prime + k_{wbg} k_{sb}^\prime \textbf{I}_{sb}^\prime \circ  \mathbb{A}_{wb}^\prime + k_{wcg} k_{sc}^\prime \textbf{I}_{sc}^\prime \circ \mathbb{A}_{wc}^\prime ~ ,
\end{eqnarray}
where $k_{sa}^\prime$ , $k_{sb}^\prime$ , and $k_{sc}^\prime$ are coefficients. $\mathbb{A}_{wa}^\prime$ , $\mathbb{A}_{wb}^\prime$ , and $\mathbb{A}_{wc}^\prime$ are the physical quantities in the subspace $\mathbb{H}_e$ . According to the multiplication table of trigintaduonions, $\textbf{I}_{sa}^\prime$ , $\textbf{I}_{sb}^\prime$ , and $\textbf{I}_{sc}^\prime$ are the physical quantities in the subspaces, $\mathbb{H}_{sa}$ , $\mathbb{H}_{sb}$ , and $\mathbb{H}_{sc}$ , respectively.

Under special circumstances, there may exist,
\begin{eqnarray}
\mathbb{A}_{wa}^\prime = \mathbb{A}_{wb}^\prime = \mathbb{A}_{wc}^\prime = \mathbb{A}_w^\prime ~ ,
\end{eqnarray}
further the trigintaduonion field potential can be reduced into,
\begin{eqnarray}
\mathbb{A} = k_{eg} \mathbb{A}_e + ( k_{wag} k_{sa}^\prime \textbf{I}_{sa}^\prime + k_{wbg} k_{sb}^\prime \textbf{I}_{sb}^\prime + k_{wcg} k_{sc}^\prime \textbf{I}_{sc}^\prime ) \circ \mathbb{A}_w^\prime ~ .
\end{eqnarray}

Apparently, the physical quantities, $\textbf{I}_{sa}^\prime$ , $\textbf{I}_{sb}^\prime$ , $\textbf{I}_{sc}^\prime$ , and $\mathbb{A}_w^\prime$ , are independent of each other. $\mathbb{A}_w^\prime$ and $\mathbb{A}_e$ are in the same subspace $\mathbb{H}_e$ . The above equation is consistent with some conclusions of field potential in the
electroweak theory \cite{ambjorn}.

In the trigintaduonion spaces, the trigintaduonion field strength may play a particularly significant role sometimes. Therefore, it is necessary to discuss several characteristics of trigintaduonion field equations, when the contributions of field strengths are quite prominent.

\section{\label{sec:level1}Prominent field strength}

The trigintaduonion field equations can be simplified further, if the trigintaduonion field strength has a particularly significant impact on other physical quantities, in the trigintaduonion spaces (Table 4).

The trigintaduonion field potential can be reduced into,
\begin{eqnarray}
\mathbb{A} = \textrm{i} \lozenge^\times \circ \mathbb{X} ~ ,
\end{eqnarray}
and the trigintaduonion field strength will be simplified into,
\begin{eqnarray}
\mathbb{F} = \lozenge \circ \mathbb{A} ~ .
\end{eqnarray}

As a result, the simplified trigintaduonion field source is written as,
\begin{eqnarray}
\mu \mathbb{S} = - ( \lozenge + k_{ff} \mathbb{F})^\ast \circ \mathbb{F} ~ ,
\end{eqnarray}
where $k_{ff} = \textrm{i} / v_0$ .

In the trigintaduonion spaces, the simplified trigintaduonion torque is,
\begin{eqnarray}
\mathbb{W} = - v_0 ( \lozenge + k_{lf} \mathbb{F} ) \circ \{ ( i \mathbb{V}^\times / v_0 ) \circ \mathbb{L} \} ~ ,
\end{eqnarray}
where $\mathbb{P} = \mu \mathbb{S} / \mu_g$ , $\mathbb{L} = ( \mathbb{R} + k_{rx} \mathbb{X} )^\times \circ \mathbb{P}$ . $k_{lf} = \textrm{i} / v_0$ .

Therefore, the simplified trigintaduonion force is,
\begin{eqnarray}
\mathbb{N} = - ( \lozenge + k_{wf} \mathbb{F} ) \circ \{ ( i \mathbb{V}^\times / v_0 ) \circ \mathbb{W} \} ~ ,
\end{eqnarray}
where $k_{wf} = \textrm{i} / v_0$ .

In some cases, the trigintaduonion field strength plays an important role, including the gravitational strength, electromagnetic strength, field strength of weak interactions, and field strength of strong interactions. These field strengths can make significant contributions to the trigintaduonion field source, linear momentum, angular momentum, torque, and force. In particular, the gravitational strengths and electromagnetic strengths have an impressive impact sometimes.

\begin{table}[h]
\caption{The quaternion operator and trigintaduonion field strength can be combined together to become one composite operator, achieving several trigintaduonion physical quantities that are related to the contributions of trigintaduonion field strengths.}
\center
\begin{tabular}{@{}ll@{}}
\hline\hline
physical quantity                      &   definition                                                                                                              \\
\hline
trigintaduonion field potential        &   $\mathbb{A} = \textrm{i} \lozenge^\times \circ \mathbb{X}$                                                              \\
trigintaduonion field strength         &   $\mathbb{F} = \lozenge \circ \mathbb{A}$                                                                                \\
trigintaduonion field source           &   $\mu \mathbb{S} = - ( \lozenge + k_{ff} \mathbb{F} )^\ast \circ \mathbb{F}$                                             \\
trigintaduonion linear momentum        &   $\mathbb{P} = \mu \mathbb{S} / \mu_g$                                                                                   \\
trigintaduonion angular momentum       &   $\mathbb{L} = ( \mathbb{R} + k_{rx} \mathbb{X} )^\times \circ \mathbb{P}$                                               \\
trigintaduonion torque                 &   $\mathbb{W} = - v_0 ( \lozenge + k_{lf} \mathbb{F} ) \circ \{ ( i \mathbb{V}^\times / v_0 ) \circ \mathbb{L} \}$        \\
trigintaduonion force                  &   $\mathbb{N} = - ( \lozenge + k_{wf} \mathbb{F} ) \circ \{ ( i \mathbb{V}^\times / v_0 ) \circ \mathbb{W} \}$            \\
\hline\hline
\end{tabular}
\end{table}

\section{\label{sec:level1}Discussions and conclusions}

The fundamental space is the extension of a fundamental field. Each fundamental field possesses one exclusive fundamental space. The gravitational field possesses an exclusive fundamental space. Similarly, the electromagnetic field possesses also another exclusive fundamental space. Due to the difference between gravitational field and electromagnetic field, the fundamental space for gravitational fields is independent of that for electromagnetic fields.

J. C. Maxwell first selected the fundamental space for electromagnetic fields as the quaternion space. Meanwhile, the fundamental space for gravitational fields can be chosen as the quaternion space also. Due to the fact that the fundamental space for gravitational fields is different from that for electromagnetic fields, the quaternion space for gravitational fields is independent of that for electromagnetic fields. Further, these two independent quaternion spaces are able to constitute an octonion space. It means that the octonion space is capable of describing simultaneously the physical characteristics of gravitational and electromagnetic fields. This point of view can be extended to the weak and strong interactions.

In the electroweak theory, the physical properties of weak interactions are quite distinct from those of electromagnetic fields. The weak interactions are composed of three fundamental fields. These three fundamental fields collectively describe the physical properties of weak interactions. Due to the independence of these three fundamental fields, it is necessary to utilize three independent quaternion spaces to describe the physical properties of weak interactions. It means that the electroweak theory involves one quaternion space for electromagnetic fields as well as three independent quaternion spaces related to weak interactions.

Further, if we want to consider the electroweak fields and gravitational fields simultaneously, it will involve a quaternion space for gravitational fields and four independent quaternion spaces related to the electroweak fields. According to the multiplication table of multidimensional spaces related to quaternion spaces, some independent quaternion spaces can be combined together to become one multidimensional space, including the octonion spaces, sedenion spaces, and trigintaduonion spaces and others. As a result, if it is going to consider the electroweak fields and gravitational fields simultaneously, we shall inevitably involve the trigintaduonion spaces in this paper. It includes one quaternion space for gravitational fields, four quaternion spaces related to the electroweak fields, and three conjugate quaternion spaces relevant to the strong interactions. This means that the strong interactions are composed of three fundamental fields. These three fundamental fields related to the strong interactions describe collectively the physical properties of strong interactions. This is consistent with the conclusions derived from the quark theory.

Compared to the cases of the octonion spaces, it is able to achieve more invariants in the trigintaduonion spaces, including the mass and electric charge, and even the anti-matter, dark matter, and dark energy and so forth. In case $\mathbb{N} = 0$ under certain conditions, one can further deduce more conservation laws, continuity equations, and equilibrium equations in the trigintaduonion spaces, including the fluid continuity equation, current continuity equation, force equilibrium equation, torque continuity equation, and precession equilibrium equation and so on.

It should be noted that the paper only discusses some simple cases of electromagnetic fields, gravitational fields, weak interactions, and strong interactions described by the algebra of trigintaduonions. But it has clearly demonstrated that the physical properties of weak and strong interactions are quite different from those of electromagnetic and gravitational interactions. It is necessary to apply three quaternion spaces independent of each other to explore the physical properties of weak interactions. Meanwhile, the physical properties of strong interactions need to be described using three mutually independent and conjugate quaternion spaces. Obviously, these three quaternion spaces related to weak interactions are completely independent of these three conjugate quaternion spaces relevant to strong interactions. In the future study, we plan to generalize the quantization characteristics in the octonion spaces to the trigintaduonion spaces. It is going to extend the adjoint fields and curved spaces in the sedenion spaces to the trigintaduonion spaces, exploring the contribution of trigintaduonion spaces on certain physical phenomena.

\section*{Acknowledgments}
The author is indebted to the anonymous referees for their valuable comments on the previous manuscripts. This project was supported partially by the National Natural Science Foundation of China under grant number 60677039.

\subsection*{Conflict of interest}
The author declares no potential conflict of interests.

\end{document}